\documentclass[preprint,tightenlines,aps,floats,nofootinbib]{revtex4}
  
\usepackage{epsf}
\usepackage{epsfig}
\usepackage{graphicx}

\newcommand{\bra}{\langle}
\newcommand{\ket}{\rangle}
\newcommand{\obar}{\overline}
\newcommand{\notE}{\ \hbox{{$E$}\kern-.60em\hbox{/}}}
\newcommand{\notp}{\ \hbox{{$p$}\kern-.43em\hbox{/}}}
\def\D0{\mbox{D\O}}

\includegraphics{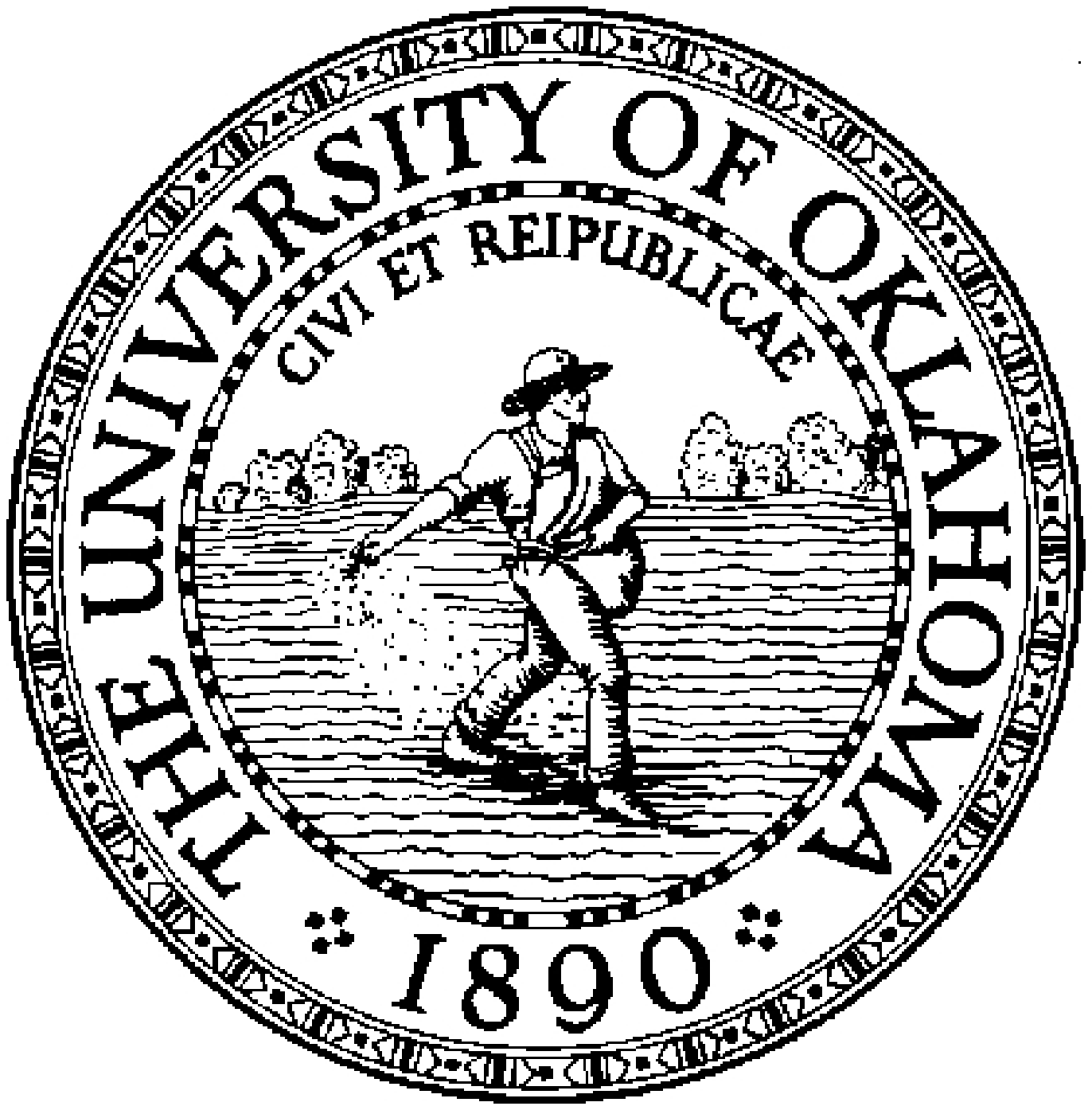}

\preprint{\font\fortssbx=cmssbx10 scaled \magstep2
\hbox to \hsize{
\hskip1.2in 
\hbox{\fortssbx The University of Oklahoma}
\hskip0.2in $\vcenter{
                      \hbox{\bf FERMILAB-PUB-06-378-T}
                      \hbox{\bf hep-ph/0610284}
                      \hbox{October 2006}}$ }
}
  
\begin{document}  

  
\title{\vspace*{0.7in}
 QCD Corrections to Higgs Pair Production \\
 in Bottom Quark Fusion}
 
\author{
Sally Dawson$^a$, Chung Kao$^{b,c}$, Yili Wang$^c$ and Peter Williams$^c$
}

\affiliation{
$^a$Department of Physics, Brookhaven National Laboratory, 
Upton, NY 11973, USA \\
$^b$Fermi National Accelerator Laboratory, P.O. Box 500, \\
Batavia, Illinois 60510, USA \\
$^c$Homer L. Dodge Department of Physics and Astronomy, 
University of Oklahoma, \\ 
Norman, Oklahoma 73019, USA 
\vspace*{.5in}}

\thispagestyle{empty}

\begin{abstract}

We present a complete next-to-leading order (NLO) calculation for 
the total cross section of inclusive Higgs pair production 
via bottom-quark fusion ($b\bar{b} \to hh$) at the CERN Large Hadron
Collider (LHC) in the Standard Model.
The NLO QCD corrections lead to less dependence on the renormalization
scale ($\mu_R$) and the factorization scale ($\mu_F$) than the
leading-order (LO) cross section, and they significantly increase the LO 
cross section. 
The rate for inclusive Higgs pair production is small in the Standard Model,
but can be large in models with enhanced couplings of the $b$ quark to
the Higgs bosons.

\end{abstract}

\pacs{PACS numbers: 14.80.Bn, 12.38.Bx, 13.85.Lg, 13.87.Ce}
%

\maketitle

\newpage

\section{Introduction}

In the Standard Model (SM), one Higgs doublet is responsible for the 
electroweak symmetry breaking (EWSB) that generates masses for gauge
bosons and fermions. A neutral Higgs boson ($h$) remains after EWSB, 
and it is the only SM elementary particle that has not been observed in
high energy experiments.  In extensions of the Standard Model, there
can be more Higgs bosons. 

One of the most important goals of the Fermilab Tevatron Run II and 
the CERN Large Hadron Collider (LHC) is to discover the Higgs bosons
or to prove their non-existence. The present lower bound on the
standard Higgs boson mass from direct searches 
at LEP2~\cite{Barate:2003sz,LEPEWWG:2005di} is $M_h > 114$~GeV. 
The electroweak precision measurements set an upper limit of 
$M_h < 166$~GeV at 95\% confidence level for the Standard Model Higgs
boson~\cite{LEPEWWG} using the recently measured top quark mass of 
$m_t = 171.4 \pm 1.2 \pm 1.8$~GeV~\cite{Top-Run2}. 
This limit increases to $199$~GeV when
the LEP2 direct search limit is included. 

The Fermilab Tevatron and the LHC will play crucial roles in Higgs searches. 
Once a candidate Higgs boson is discovered, it will be necessary to 
determine the Higgs couplings and spin to see if the Higgs candidate
has the properties of the SM Higgs boson.  One of the most difficult
properties to measure is the trilinear self coupling of 
a Higgs boson~\cite{
Boudjema:1995cb,Djouadi:1999rc,Muhlleitner:2003me,Baur:2003gp,Moretti:2004wa}.
The high energy and high luminosity at the LHC might provide 
opportunities to detect a pair of Higgs bosons as well as a discovery
channel to measure the trilinear Higgs couplings in the SM and 
in models with more Higgs bosons.
In the Standard Model, gluon fusion is the dominant process to produce 
a pair of Higgs bosons via triangle and box diagrams with internal top 
quarks and bottom quarks~\cite{
Dicus:1987ic,Glover:1987nx,Dawson:1998py,BarrientosBendezu:2001di}. 
Bottom quark fusion can also produce Higgs pairs at 
a lower rate.

At tree level, the physical production mechanism for a Higgs boson
pair in association with $b$ quarks is $gg\to b {\overline b} hh$.  
This process contains a large collinear logarithm from the
gluon splitting into a collinear $b\bar{b}$ pair, 
$\Lambda \equiv \ln( M_h/m_b )$.  
These logarithms can be resummed by using a perturbatively defined 
$b$ quark parton distribution function (PDF) which is inherently 
$O(\alpha_s\Lambda)$~\cite{
Olness:1987:ae,Barnett:1987jw,Stelzer:1998ni,Dicus:1998hs}. 
In this approach, the ordering of perturbation theory is changed to be
an expansion in ${\cal O}(\alpha_s)$ and $\Lambda^{-1}$.

When using a scheme with $b$ quark PDFs for Higgs pair production 
in association with $b$ quarks, the leading order (LO) process becomes 
$b {\overline b}\to hh$, and we compute the NLO cross section 
with $O(\alpha_s)$ and $O(1/\Lambda)$ corrections to this process.  
The subprocess $bg\to b hh$ is a correction of $O(1/\Lambda)$
to the lowest order process, while $gg\to b {\overline b} hh$
is $O(1/\Lambda^2)$.

The rate for Higgs pair production at the LHC is small in the SM.  However,
it can become significant in models in which the Higgs coupling to
the bottom quark is enhanced~\cite{Jin:2005gw}. 
In two Higgs doublet models with Model II type of Yukawa
interactions (including the minimal supersymmetric standard model (MSSM)), 
the ratio of the Higgs vacuum expectation values 
($\tan\beta \equiv v_2/v_1$) is an important parameter. 
A large value of $\tan\beta$ greatly enhances the Higgs coupling 
with bottom quarks and makes bottom quark fusion become the largest
production mechanism for producing a Higgs boson at the LHC.  
The D0 experiment has placed limits on single Higgs production in 
association with $b$ quarks for large values of $\tan\beta$~\cite{ichep}.

In this paper, we present the complete next-to-leading order (NLO) 
calculations to the production of Higgs pairs via bottom quark fusion 
in the Standard Model.  We compute a consistent set of
$O(\alpha_s)$ and $O(1/\Lambda)$ corrections.
In a future paper, we will present results for double Higgs production
at NLO in the MSSM~\cite{future}.

The theoretical prediction depends on the number of $b$ quarks tagged.
Here, we consider only  inclusive processes in which there
are no tagged $b$ quarks.
We apply the two cutoff phase space slicing 
method~\cite{Ohnemus:1990za,Harris:2001sx} to calculate corrections
from real gluon emission. Two arbitrary small parameters are
introduced to split the phase space in real gluon emission into soft, 
hard/collinear, and hard/non-collinear regions. 
The production in the hard/non-collinear region is finite and can be 
calculated numerically. 
Divergences are isolated into soft and hard/collinear regions. 
The soft (infrared) singularities cancel with the infrared
singularities in the virtual corrections. The collinear singularities
are absorbed into the initial parton distribution functions. 

In section II, we compute the leading-order cross section for 
$pp \to hh +X$  via $b\bar{b} \to hh$. 
In section III, we provide a complete  next-leading order (NLO)
calculation for $hh$ production. At the parton level, we compute
one-loop virtual corrections and real gluon emission corrections to 
$b\bar{b}  \to hh$ and the ${\cal O}(1/\Lambda)$ subleading process,
$bg\to b hh$. 
The associated production,  $g g \to b\bar{b} hh$, is finite and is
a subleading correction of ${\cal O}(1/\Lambda^2)$
to the inclusive rate for $pp \to hh +X$.
We use MadGraph~\cite{MadGraph} and HELAS~\cite{HELAS} to compute 
its LO production rate with a finite mass of the bottom quark. 
Numerical results are given in Section IV and 
conclusions are drawn in section V. In addition, there is an 
appendix to present formulas for the $b$ quark running mass.

\section{Leading-order Cross Section for $b\bar{b} \to hh$}

\begin{figure}[htb]
\centering\leavevmode
\epsfxsize=6in
\epsfbox{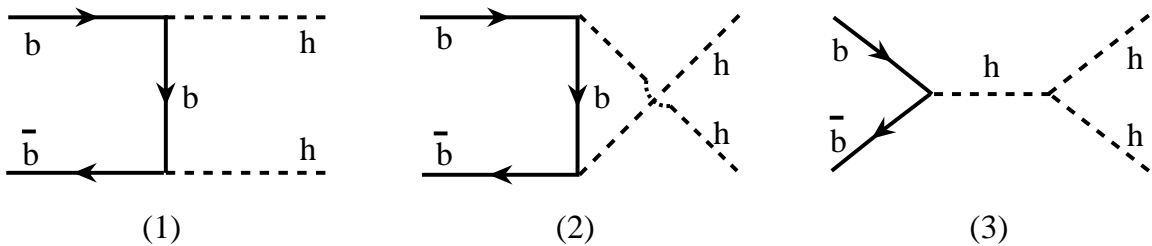}
\caption[]{The lowest order Feynman diagrams for $b\bar{b} \to hh$.}
\label{fig:tree}
\end{figure}

The leading order (LO) inclusive cross section for $pp \to hh +X$ 
via  $b\bar{b} \to hh $ is evaluated with 
\begin{eqnarray}
\sigma_{LO} = \int dx_1 dx_2 
              \left[ b(x_1) \bar{b}(x_2) +\bar{b}(x_1) b(x_2) \right]
              \hat{\sigma}_{LO}(s,t,u)(b\bar{b} \to hh)
\label{sigma:lo}
\end{eqnarray}
where $b(x)$ and $\bar{b}(x)$ are the LO parton distribution functions for 
bottom quarks in  the proton, $\hat{\sigma}_{LO}(s,t,u)$ is the parton
level cross section for $b\bar{b} \to hh $ and $s,t,u$ are the
Mandelstam variables. Fig.~\ref{fig:tree} shows the tree level Feynman 
diagrams for $b\bar{b} \to hh $.


We assign momenta to the initial and the final state particles with 
$$b(p_1)\bar{b}(p_2) \to h(p_3) h(p_4) $$ and $p_1 + p_2 = p_3 + p_4$. 
We take the $b\bar{b}h$ and $hhh$ couplings to be
$-i\frac{m_b}{v}C_{bh}$ and $-i3\frac{M_h^2}{v}C_{hhh}$, respectively, 
with $v =$ the Higgs vacuum expectation value $\simeq 246$ GeV, 
$C_{bbh} = 1$ and $C_{hhh} = 1$ in the Standard Model. 
We evaluate the bottom quark mass in the $h b\bar{b}$ Yukawa coupling 
by using the $\overline{MS}$ mass, $\bar{m}_b(\mu)$ (defined in Appendix A), 
for a two-loop heavy quark running mass~\cite{Vermaseren:1997fq}
with $m_b({\rm pole}) = 4.7$ GeV and the NLO evolution of 
the strong coupling~\cite{Marciano:1983pj}.

In addition,the Mandelstam variables are defined as 
\begin{eqnarray*}
s & = & (p_1 + p_2)^2 \nonumber \\
t & = & (p_1 - p_3)^2  \nonumber \\
u & = & (p_2 - p_3)^2 \quad .
\end{eqnarray*}

Following the simplified 
ACOT prescription~\cite{Aivazis:1993pi,Collins:1998rz,Kramer:2000hn}, 
we take $m_b = 0$ everywhere except in the Yukawa couplings. 
Then the tree level amplitudes of the $s, t$ and $u$ channels are:
\begin{eqnarray*}
M_s^0
 & \equiv & \hat{M^0_s} \delta_{ji}
 = -\frac{3  C_{bh} C_{hhh} \bar{m}_b(\mu) M_h^2}
         {v^2 \left(s -M_h^2 + i M_h \Gamma_h\right)}
    \delta_{ji}\bar{v}(p_2) u(p_1) \nonumber \\
M_t^0
 & \equiv & \hat{M^0_t} \delta_{ji}
 = \frac{ C_{bh}^2 \bar{m}^2_b(\mu)}{v^2 t}
   \delta_{ji}\bar{v}(p_2)\notp_3 u(p_1) \nonumber \\
M_u^0
 & \equiv & \hat{M^0_u} \delta_{ji}
 = -\frac{ C_{bh}^2 \bar{m}^2_b(\mu)}{v^2 u}
   \delta_{ji}\bar{v}(p_2)\notp_3 u(p_1)
\end{eqnarray*}
where $i$ and $j$ are color indices for initial $b$ and $\bar{b}$
quarks. The $\mu$ parameter is an arbitrary mass, which is introduced 
such that both  the renormalized strong coupling and Yukawa coupling are 
dimensionless in $N$ dimensions.  
The corresponding spin- and color-averaged matrix elements squared 
including interference terms are
\begin{eqnarray*}
&&\bra |M_s^0|^2 \ket
  =  \frac{3}{2}\left(\frac{\bar{m}_b^2(\mu)}{v^2}\right) C_{bh}^2 C_{hhh}^2 \frac{s}{v^2}\left|\frac{M_h^2}{s - M_h^2 + i \Gamma_h M_h}\right|^2 \nonumber \\
&&\bra |M_t^0|^2 \ket
  =  \frac{1}{6} \left(\frac{\bar{m}_b^4(\mu)}{v^4}\right) C_{bh}^4 \left(\frac{u}{t} - \frac{M_h^4}{t^2}\right) \nonumber \\
&&\bra |M_u^0|^2 \ket
  =  \frac{1}{6} \left(\frac{\bar{m}_b^4(\mu)}{v^4}\right) C_{bh}^4 \left(\frac{t}{u} - \frac{M_h^4}{u^2}\right) \nonumber \\
&& \bra Re(M_t^0\bar{M_u^0}) \ket
  = -\frac{1}{6} \left(\frac{\bar{m}_b^4(\mu)}{v^4}\right)
     C_{bh}^4 \left(1 - \frac{M_h^4}{t u}\right) \quad ,
\end{eqnarray*}
where $\Gamma_h$ is the decay width of the Higgs boson.

Summing the above terms, we obtain the total matrix element squared
\begin{eqnarray*}
\bra |M_0|^2 \ket 
& = & \bra |M_s^0|^2 \ket +\bra |M_t^0|^2 \ket +\bra |M_u^0|^2 \ket
     +2 \bra Re(M_t^0\obar{M}_u^0) \ket \\
& = & \frac{1}{6}\left(\frac{\bar{m}_b^2(\mu)}{v^2}\right) C^2_{bh}
      \left[ 9\frac{s}{v^2} C_{hhh}^2 
      \left|\frac{M_h^2}{s -M_h^2 + i M_h \Gamma_h}\right|^2 \right. \\
&   &\left. +\left(\frac{\bar{m}_b^2(\mu)}{v^2}\right) C^2_{bh}
     \left(1 - \frac{M^4_h}{u t}\right)\frac{(u - t)^2}{u t} \right] \quad .
\end{eqnarray*}
The parton level cross section for inclusive  $b\bar{b} \to hh$
production becomes 
\begin{eqnarray}
\hat{\sigma}_{\rm LO} =
 \int\frac{1}{2 s}\frac{1}{2}\bra |M_0|^2 \ket
 d \Phi_2(b\bar{b} \to hh)
\end{eqnarray}
where $d\Phi_2(b\bar{b} \to hh)$ denotes the two-body phase space
factor and a factor of $1/2$ comes from identical particles 
in the final state.

\section{Next-To-Leading Order Corrections for $b\bar{b} \to hh$}

To determine the next-to-leading order (NLO) corrections for Higgs
boson pair production in bottom quark fusion, 
we evaluate the cross section for 
(a) the leading-order subprocess ($b\bar{b} \to hh$),
(b) the $\alpha_s$ corrections, and  
(c) the $1/\Lambda$ corrections ($bg \to b hh$), 
where $\Lambda \equiv \ln(M_h/m_b)$. 
The order $\alpha_s$ corrections have contributions from 
one-loop diagrams with virtual gluons ($b\bar{b} \to hh$) and 
tree-level real gluon emission ($b\bar{b} \to hhg$). 
The NLO correction contains
both the $\alpha_s$ and the $1/\Lambda$ corrections.

We write the parton level NLO cross section as
\begin{eqnarray}
  \hat{\sigma}_{\rm NLO}(x_1,x_2, \mu) 
& = & \hat{\sigma}_{\rm LO}(x_1,x_2, \mu) 
     +\delta\hat{\sigma}_{NLO}(x_1, x_2, \mu) \nonumber \\
  \delta\hat{\sigma}_{\rm NLO}(x_1,x_2, \mu) 
& = & \delta\hat{\sigma}_{\alpha_s}(x_1, x_2, \mu)
     +\delta\hat{\sigma}_{1/\Lambda}(x_1, x_2, \mu)
\label{sigma:pnlo}
\end{eqnarray} 
where $\hat{\sigma}_{\rm LO}(x_1, x_2, \mu)$ is the leading order
(Born) cross section and $\delta\hat{\sigma}_{\rm NLO}(x_1, x_2, \mu)$ 
is the next-to-leading order correction to the Born cross section, 
$x_{1,2}$ are the  momentum fractions of the partons, $x_1 x_2 = \hat{s}/S$,
$S$ is the  center of mass energy of the hadrons and $\mu = \mu_R$ is the
renormalization scale. 

The ${\cal O}(\alpha_s)$ correction includes contributions from both 
virtual and real gluon emission:
\begin{eqnarray}
\delta\hat{\sigma}_{\alpha_s}(x_1,x_2, \mu ) 
=  \delta\hat{\sigma}_{v}(x_1, x_2, \mu) 
  +\delta\hat{\sigma}_{r}(x_1, x_2, \mu)
\label{sigma:vr}
\end{eqnarray} 
where $\delta\hat{\sigma}_{v}(x_1,x_2,\mu)$ and 
$\delta\hat{\sigma}_{r}(x_1,x_2,\mu)$ 
represent virtual and real gluon emission NLO corrections to 
$b\bar{b} \to hh$.

\subsection{Corrections with virtual gluons}

\begin{figure}[htb]
\centering\leavevmode
\epsfxsize=6in
\epsfbox{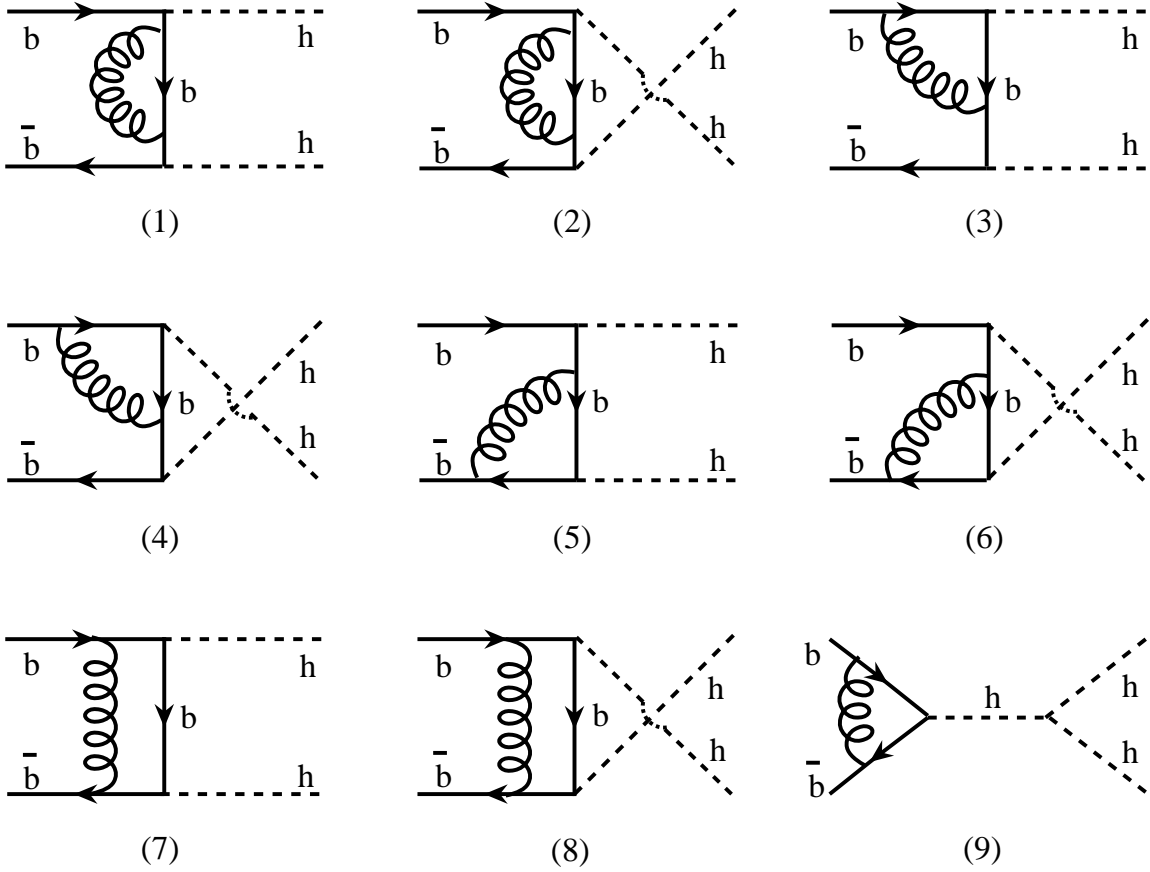}
\caption[]{One-loop virtual corrections to $b\bar{b} \to hh$.}
\label{fig:oneloop}
\end{figure}

The one-loop diagrams for the ${\cal O}(\alpha_s)$ corrections to 
$b\bar{b} \to hh $ are shown in Fig.~\ref{fig:oneloop}. 
The $\alpha_s$ corrections involving virtual gluons 
are evaluated as the interferences between Born diagrams  
(Fig.~\ref{fig:tree}) and one-loop virtual diagrams (Fig.~\ref{fig:oneloop}). 
\begin{eqnarray*}
  |M_v|^2
 = M_0\obar{M}_{loop} +M_{loop}\obar{M}_0 
 = 2 Re(M_{loop}\obar{M}_0) \;.
\end{eqnarray*}


We evaluate the amplitudes of  the one-loop diagrams by applying dimensional 
regularization in $N$ dimensions with $N \equiv 4-2\epsilon$.
The virtual diagrams are computed analytically and all tensor 
integrals are reduced to linear combinations of one-loop 
scalar functions. All amplitudes for the virtual corrections can be 
reduced to Born amplitudes multiplied by
coefficients $X_i$. 
We find,
\begin{eqnarray}
2 Re( M_{loop}\obar{M_0})
 = 2 C_F g_s^2 Re\{ \left[ X_s |M_s^0|^2 + X_t |M_t^0|^2 + X_u |M_u^0|^2
                   +(X_t + X_u) Re( M_t^0\obar{M_u^0} ) \right] \}
\end{eqnarray}
where $C_F = 4/3$ and
\begin{eqnarray*}
X_s & \equiv & X_9 \nonumber \\
X_t & \equiv & X_1 + X_3 + X_5 + X_7  \nonumber \\
X_u & \equiv & X_2 + X_4 + X_6 + X_8 \nonumber
\end{eqnarray*}
The coefficients are,
\begin{eqnarray}
X_1 & = & -A \left(\frac{1}{\epsilon} + 1-\ln(-t) \right)  \nonumber \\
X_2 & = & -A \left(\frac{1}{\epsilon} + 1-\ln(-u) \right)  \nonumber \\
X_3 & = & X_5 = 
 2A \left\{ (-t)^{-\epsilon} \left(\frac{1}{\epsilon} + 2 \right) - \left(\frac{M^2_h}{M^2_h-t}\right) \left[\frac{1}{\epsilon} \ln\left(\frac{-t}{M^2_h}\right) + \frac{1}{2} \ln^2(M^2_h)-\frac{1}{2}\ln^2(-t)-\frac{\pi^2}{2} \right] -1 \right\} \nonumber \\
X_4 & = & X_6 =
 2A \left\{ (-u)^{-\epsilon} \left(\frac{1}{\epsilon} + 2 \right) - \left(\frac{M^2_h}{M^2_h-u}\right) \left[\frac{1}{\epsilon} \ln\left(\frac{-u}{M^2_h}\right) + \frac{1}{2} \ln^2(M^2_h)-\frac{1}{2}\ln^2(-u)-\frac{\pi^2}{2} \right] -1 \right\} \nonumber \\
X_7 & = &  2A \left\{- \frac{1}{\epsilon^2} + \frac{1}{\epsilon}\left[\frac{2M^2_h}{M^2_h - t} \ln\left(\frac{-t}{M^2_h}\right) + \ln(s)\right] + F(t) \right\} \nonumber \\
X_8 & = &  2A \left\{- \frac{1}{\epsilon^2} + \frac{1}{\epsilon}\left[\frac{2M^2_h}{M^2_h - u} \ln\left(\frac{-u}{M^2_h}\right) + \ln(s)\right] + F(u) \right\}\nonumber \\
X_9 & = & -2A \left[\frac{1}{\epsilon^2}  -
   \frac{1}{\epsilon}\ln(s) + \frac{1}{2}\ln^2(s) + 1
  -\frac{2\pi^2}{3}\right] \quad ,
\end{eqnarray}
where $A$ is a normalization factor
\begin{eqnarray*}
A = \frac{1}{16 \pi^2} \Gamma(1+\epsilon) (4\pi\mu^2)^\epsilon.
\end{eqnarray*}
The function $F(x)$ is defined as 
\begin{eqnarray}
F(x) & = &\frac{-x}{M^2_h -x }\left[-\ln^2(M^2_h) + \ln^2(-x) +\pi^2\right] \nonumber \\
& + & \frac{x}{s\beta_h}\left[\ln\left(\frac{M^2_h}{s}\right)\ln \left(\frac{1-\beta_h}{1 + \beta_h} \right)+ 2 Li_2\left(\frac{1+\beta_h}{2}\right) - 2 Li_2\left(\frac{1-\beta_h}{2}\right) \right]  \nonumber \\
&-&   2 \ln^2\left(\frac{-x \sqrt{s}}{M_h^2}\right)-4Li_2\left(1-\frac{x}{M^2_h}\right) + \frac{2\pi^2}{3} 
\end{eqnarray}
\noindent
where  $\beta_h = \sqrt{1-4M^2_h/s}$ and $Li_2$ is the dilogarithm 
or the Spence function~\cite{Spence}.

The virtual corrections contain both ultraviolet (UV) and infrared (IR) 
divergences. In the $\overline{MS}$ scheme, the $b$ quark Yukawa
coupling is renormalized with the counter term~\cite{Braaten:1980yq} 
\begin{eqnarray*}
\frac{\delta m_b}{m_b} = - A \frac{16 \pi \alpha_s}{\epsilon} \quad .
\end{eqnarray*}
This counter term contributes to the total matrix element squared by
\begin{eqnarray*}
 |M_{\rm CT}|^2
 & = & 2 \frac{\delta m_b}{m_b} |M_s^0|^2
      +4 \frac{\delta m_b}{m_b}
       \left( |M_t^0|^2 +|M_u^0|^2 +2 Re(M_t^0\obar{M_u^0}) \right) \\
 & = & -\frac{32 A \pi\alpha_s}{\epsilon}\left[
      |M_s^0|^2 + 2 (|M_t^0|^2+|M_u^0|^2+2 Re M_t^0\obar{M_u^0} ) \right]
     \quad .
\end{eqnarray*}

Summing over all relevant contributions, we obtain the following
expression for the one-loop virtual corrections
\begin{eqnarray}
|M_v|^2
& = & 2 Re( M_{loop}\obar{M_0} ) + |M_{\rm CT}|^2 \nonumber \\
& = & |M_0|^2 \left\{ A \frac{64\pi\alpha_s}{3}
      \left[ -\frac{1}{\epsilon^2} +\frac{1}{\epsilon}\ln(s)
             -\frac{3}{2\epsilon} \right] \right\}
             +A\frac{64 \pi \alpha_s}{3} |M_D|^2\quad ,
\label{eq:v}
\end{eqnarray}
where $|M_D|^2$ contains the 
finite terms: 
\begin{eqnarray}
 |M_D|^2 & = & \left[-\frac{1}{2}\ln^2(s) + \frac{2 \pi^2}{3} -1 \right] |{M_s^0}|^2  \nonumber \\
& +&  \frac{M_h^2}{M_h^2 -t} \left[- \ln^2(M_h^2) + \ln^2(-t) + \pi^2 \right]|M_t^0|^2 + \left[F(t) + \frac{3}{2} -\frac{3}{2} \ln(-t)\right] |M_t^0|^2 \nonumber \\
& + &  \frac{M_h^2}{M_h^2 -u} \left[-\ln^2(M_h^2) + \ln^2(-u) +\pi^2 \right] |M_u^0|^2 +\left[F(u) +\frac{3}{2} -  \frac{3}{2} \ln(-u)\right] |{M_u^0}|^2 \nonumber \\
&+& \frac{M_h^4}{(M_h^2-t)(M_h^2 -u)}\left[\ln^2(-t) + \ln^2(-u) +F(t) + F(u)\right. \nonumber \\
& - & \left. \frac{3}{2}\ln(-t) -\frac{3}{2}\ln(-u) + 2 \pi^2 + 3\right] Re({M_t^0} \overline{M_u^0})  \nonumber \\
& + & \frac{M_h^2}{(M_h^2-t)(M_h^2 -u)} \left\{
 -s \ln^2(M_h^2) - u \ln^2(-t) - t \ln^2(-u)\right.\nonumber \\
& - & \left. \left[F(t) + F(u) - \frac{3}{2} \ln(-t) - \frac{3}{2} \ln(-u) + \pi^2 + 3\right](t+u) \right\} Re({M_t^0}\overline{ M_u^0})  \nonumber \\
&  + & \frac{t u}{(M_h^2-t)(M_h^2 -u)} \left[F(t) + F(u) + 3 - \frac{3}{2} \ln(-t) -\frac{3}{2} \ln(-u)\right] Re({M_t^0}\overline{M_u^0}) 
\quad .
\end{eqnarray}
The divergences left in Eq.~\ref{eq:v} are infrared divergences
which are canceled by the infrared divergences in the
real gluon emission corrections
discussed in the next subsection.

\subsection{Real gluon emission}

\begin{figure}[htb]
\epsfxsize=4.5in
\begin{center}
\epsfbox{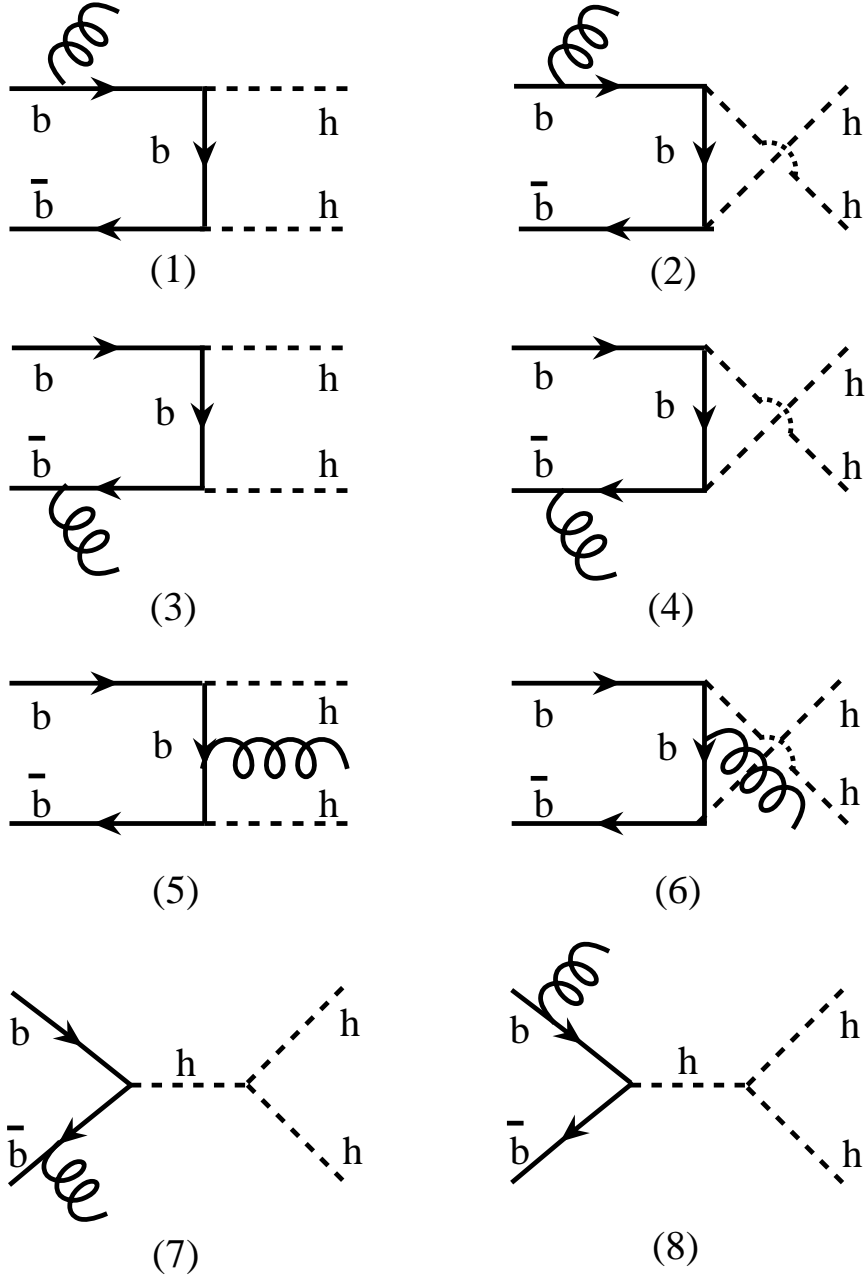}
\end{center}
\caption[]{Feynman diagrams for Higgs pair production in bottom quark
fusion with real gluon emission ($b\bar{b} \to hhg$).}
\label{fig:real}
\end{figure}

The Feynman diagrams for real gluon emission 
($b \bar{b} \to h h g$)  are shown in Fig.~\ref{fig:real}. 
We assign the momentum 
as $$b(p_1) \bar{b}(p_2) \to h(p_3) h(p_4) g(p_g).$$ 
The real gluon emission corrections have infrared (IR) and collinear singularities.  The infrared singularities cancel  the virtual infrared singularities in Eq.~\ref{eq:v} and  the collinear singularities are absorbed into parton distribution functions. We apply the  two cut-off phase space slicing 
(PSS) method~\cite{Ohnemus:1990za,Harris:2001sx} 
to isolate these singularities in different regions of phase space. 

We introduce two small cutoffs to split the phase space in the real gluon
emission process. First, we introduce a soft cutoff ($\delta_s$) 
to separate the phase space of the process $b\bar{b} \to h h g$ 
into $soft$ and $hard$ regions according to the emitted gluon energy. 
The soft region is the region where the radiated gluon energy satisfies 
$E_g < \delta_s \frac{\sqrt{s}}{2}$, while the hard region is the
region where the gluon energy satisfies,
 $ E_g \geq \delta_s \frac{\sqrt{s}}{2}$. 
Then the contributions from real gluon emission can be decomposed as 
\begin{eqnarray}
\hat{\sigma}_{r} = \hat{\sigma}_{soft} + \hat{\sigma}_{hard} \;.
\end{eqnarray}


Although the energy of the emitted gluon in the hard region is above the 
threshold, there still exist singularities when the emitted gluon 
is parallel to one of the initial bottom quarks.
We introduce a collinear cutoff ($\delta_c$) to isolate this collinear 
singularity. The phase space in the hard region is decomposed into 
hard/collinear and hard/non-collinear regions. In the hard/collinear 
portion, the gluon is emitted within an angle satisfying
\begin{eqnarray}
\frac{2p_1\cdot p_g}{E_g \sqrt{s}} < \delta_c \;\;\; or \;\;\; 
\frac{2p_2\cdot p_g}{E_g \sqrt{s}} < \delta_c \quad . 
\end{eqnarray}

The parton level cross section in the hard region is split into
hard/collinear and hard/noncollinear regions,
\begin{eqnarray}
\hat{\sigma}_{hard} = \hat{\sigma}_{hard/c} + \hat{\sigma}_{hard/nc}
\quad ,
\end{eqnarray}
where $\hat{\sigma}_{hard/c}$ is obtained by integrating over the
hard/collinear region of the gluon phase space. It includes the 
collinear singularities and can be evaluated analytically in $N$ dimensions. 
The hard non-collinear cross section ($\hat{\sigma}_{hard/nc}$) is finite. 
We have calculated $\hat{\sigma}_{hard/c}$ and $\hat{\sigma}_{hard/nc}$ 
numerically  with a collinear cutoff ($\delta_c$). 
The analytic matrix elements squared are used in our calculations 
for $\alpha_s$ and $1/\Lambda$ corrections.
Since $\delta_s$ and $\delta_c$ are arbitrary cut-offs, the dependence 
of the cross section on $\delta_s$ and $\delta_c$ is
not physical and cancels in the total NLO cross section.
  
\subsubsection{Soft gluons}

The soft gluon emission amplitudes for the diagrams in Fig.~\ref{fig:real} 
are obtained by setting the gluon momentum $p_g$ to zero everywhere 
except in the denominators that are 
singular as $p_g \to 0$. The soft gluon
emission corrections are
\begin{eqnarray}
M_{soft} = g_s^2 T^a_{ji}\left(\frac{p_2^\mu}{p_2 \cdot p_g}
          -\frac{p_1^\mu}{p_1 \cdot p_g}\right) (\hat{M}_s^0 
          +\hat{M}_t^0 + \hat{M}_u^0) \;.
\end{eqnarray}

There is a logarithmic divergence in the integral of the matrix
element squared over the  soft three-body phase space. 
The three-body phase space in the soft gluon emission approximation
is,
\begin{eqnarray}
d\Phi_3|_{soft}
& = & \left[\frac{d^{N-1}p_3}{2E_3 (2\pi)^{N-1}}
      \frac{d^{N-1}p_4}{2E_4 (2\pi)^{N-1}} (2\pi)^N
      \delta^N(p_1+p_2-p_3-p_4)
      \right] \frac{d^{N-1}p_g}{2E_g (2\pi)^{N-1}} \nonumber \\
& = & (d\Phi_2)( d\Phi_g|_{soft})
\end{eqnarray}
where $d\Phi_2$ is the two-body  phase space factor and $d\Phi_g|_{soft}$ is 
the soft gluon phase space. In the center of mass frame of the incoming
partons,
\begin{eqnarray}
d\Phi_g|_{soft}
& = & \frac{d^{N-1}p_g}{(2\pi)^{N-1} 2 E_g} \nonumber \\
& = & \frac{\Gamma(1-\epsilon)}{\Gamma(1-2\epsilon)}
      \frac{\pi^\epsilon}{(2\pi)^3}
      \int_{0}^{\frac{\sqrt{s}}{2}\delta_s} dE_g E_g^{1-2\epsilon}
      \int_0^\pi d\theta_1 sin^{1-2\epsilon}\theta_1
      \int_0^\pi d\theta_2 sin^{-2\epsilon}\theta_2 \;.
\end{eqnarray}
Together with the matrix element in the corresponding soft
approximation, this integral can be evaluated analytically, 
\begin{eqnarray}
\bra |M^\prime_{soft}|^2 \ket
& \equiv & \int d\Phi_g|_{soft} \bra |M_{soft}|^2 \ket \nonumber \\
& = & 4 \pi \alpha_s \bra |M_0|^2 \ket C_F
      \frac{1}{4 \pi^2 \epsilon^2}
      \left(\frac{4 \pi \mu^2}{\delta_s^2 s}\right)^\epsilon 
      \frac{\Gamma(1-\epsilon)}{\Gamma(1-2\epsilon)} \nonumber \\
& = & \bra |M_0|^2 \ket A \frac{64 \pi \alpha_s}{3} 
      \left[ \frac{1}{\epsilon^2} -\frac{1}{\epsilon} \ln(\delta_s^2)
            -\frac{1}{\epsilon}\ln(s) +\frac{1}{2} \ln^2(s \delta_s^2) 
            -\frac{\pi^2}{3} \right] \quad .
\label{soft}
\end{eqnarray}

The divergences in Eq.~\ref{soft} cancel the IR singularities 
in  Eq.~\ref{eq:v} of the virtual corrections. Adding Eq.~\ref{eq:v} and 
Eq.~\ref{soft} together we obtain
\begin{eqnarray}
\bra |M_v|^2 \ket +\bra |M'_{soft}|^2 \ket 
& = & \bra |M_0|^2 \ket 
      \left\{ A \frac{64 \pi \alpha_s}{3} 
             \left(-\frac{1}{\epsilon}\right)
             \left[ \ln(\delta_s^2) + \frac{3}{2} \right] \right. \nonumber \\
&  &\left. +A \frac{64 \pi \alpha_s}{3} 
     \left[\frac{1}{2} \ln^2(s \delta_s^2)
       -\frac{\pi^2}{3}\right]\right\}
    +A \frac{64 \pi \alpha_s}{3} |M_D|^2 \;. \nonumber \\
\label{soft-1}
\end{eqnarray}
However, this equation still has a collinear singularity, 
which can be absorbed into the parton distribution functions. 
We discuss this collinear singularity in the next subsection.
 
\subsubsection{Hard gluons}

In the hard/collinear region, the hard gluon is emitted collinearly
to one of the initial partons. The phase space is greatly simplified 
in the collinear limit. The initial-state $b$ quark splits into a hard 
parton $b'$ and a collinear hard gluon $g$ by $b \to b^\prime g$ 
with approximately $p_{b^\prime} = z p_b$ and $p_g = (1-z) p_b$. 
In the hard/collinear limit, the 
matrix element squared for $b\bar{b} \to hhg$ factorizes into 
the Born matrix element squared and the Altarelli-Parisi splitting
function for $b \to b^\prime g$
\begin{eqnarray}
  \bra \sum |M_{hard/c}|^2 \ket (b\bar{b}\to hhg) \to
  (4\pi\alpha_s) \bra \sum |M_0|^2 \ket 
  \frac{-2 P_{bb}(z,\epsilon)}{z (p_1 - p_g)^2} (\mu^2)^\epsilon
 +(1 \leftrightarrow 2) \quad ,
\label{col}
\end{eqnarray}
where $P_{bb}$ is the Altarelli-Parisi splitting function for 
$b \to b^\prime g$ at the lowest order
\begin{eqnarray}
P_{bb}(z, \epsilon) & = & C_F \left[\frac{1+z^2}{1-z} -\epsilon (1-z)\right]\nonumber \\
&=& P_{bb}(z) + \epsilon P^\prime_{bb}(z) \;.
\end{eqnarray}
  
The process can be factorized in two steps. First, one incoming $b$
quark radiates a hard gluon and becomes $b^\prime$. Then this
$b^\prime$ collides with another incoming $b$ quark to produce two 
Higgs bosons by  $b^\prime \bar{b} \to h h$. 
The hard/collinear phase space is,
\begin{eqnarray}
d\Phi_3\mid_{hard/c}= d\Phi_2(z p_1 +
p_2 \to p_3 + p_4 ) \frac{d^{N-1}p_g}{(2\pi)^{N-1} 2 E_g} \;.
\end{eqnarray}
To carry out the integration over $d^{N-1}p_g$ in the collinear approximation, 
we introduce a new variable, $s_{bg} = 2 p_1 \cdot p_g$, which is 
constrained by $0 \leq s_{bg} \leq \frac{s}{2} (1-z) \delta_c$.
The hard/collinear phase space of the gluon becomes,
\begin{eqnarray}
  \frac{d^{N-1}p_g}{(2\pi)^{N-1} 2 E_g}\mid_{hard/c}
= \frac{(4\pi)^\epsilon}{16 \pi^2}\frac{1}{\Gamma(1-\epsilon)} 
  dz ds_{bg} [(1-z)s_{bg}]^{-\epsilon} \quad .
\label{gluon/ph}
\end{eqnarray}

The $ds_{bg}$ integral can be evaluated to find 
the cross section in the hard/collinear region,
\begin{eqnarray}
\sigma_{hard/c}
& = & \int dx_1 dx_2 \; \bar{b}(x_2) 
      \frac{\alpha_s}{2\pi}\frac{\Gamma(1-\epsilon)}{\Gamma(1-2\epsilon)} 
      \left(\frac{4\pi\mu^2}{s}\right)^\epsilon (-\frac{1}{\epsilon}) 
      \delta_c^{-\epsilon}
      \int_{x_1}^{1-\delta_s} \frac{dz}{z} b(x_1/z) P_{bb}(z,\epsilon) 
      \hat{\sigma}_{\rm LO} \nonumber \\
&   & \times \left[\frac{(1-z)^2}{2}\right]^{-\epsilon}
      +( b \leftrightarrow \bar{b} ) +( 1 \leftrightarrow 2 ) \quad .
\label{hard/c}
\end{eqnarray}

This equation has collinear divergences which we remove 
by absorbing them into bare parton distribution functions.
We introduce a modified (NLO) parton distribution function 
at the factorization scale $\mu_F$ in the $\overline{MS}$ scheme:
\begin{eqnarray}
b(x)& = &b(x,\mu_F) \left\{1 + \frac{2\alpha_s}{3\pi}(4\pi)^\epsilon\Gamma(1+\epsilon)\left(\frac{1}{\epsilon}\right)\left[\ln(\delta_s^2) + \frac{3}{2}\right]\right\}\nonumber \\
& +
&\frac{\alpha_s}{2\pi}\frac{\Gamma(1-\epsilon)}{\Gamma(1-2\epsilon)}
(4\pi)^\epsilon \left(\frac{1}{\epsilon}\right)
\int_{x}^{1-\delta_s}P_{bb}(z) \frac{dz}{z} b(x/z) \;.
\label{pdf}
\end{eqnarray}

Replacing $b(x)$ in the lowest order hadronic cross section by $b(x,\mu_F)$
and dropping terms higher than ${\cal O}(\alpha_s)$,
we obtain the Born cross section that is proportional to $\alpha_s$  
\begin{eqnarray}
\sigma_{\rm Born}
& = & \int dx_1 dx_2 b(x_1,\mu_F)\bar{b}(x_2,\mu_F) 
      \hat{\sigma}_{LO}(x_1,x_2,\mu_R) \nonumber \\
&   &+\int dx_1 dx_2 b(x_1,\mu_F)\bar{b}(x_2,\mu_F) 
      \hat{\sigma}_{LO}(x_1,x_2,\mu_R) 
      \left\{ \frac{4\alpha_s}{3\pi}(4\pi)^\epsilon\Gamma(1+\epsilon)
              \left(\frac{1}{\epsilon}\right)\left[\ln(\delta_s^2)
             +\frac{3}{2}\right] \right\} \nonumber \\
&   &+\left\{\int dx_1 dx_2 \bar{b}(x_2,\mu_F) 
       \hat{\sigma}_{LO}(x_1,x_2,\mu_R)\frac{\alpha_s}{2\pi}
       \frac{\Gamma(1-\epsilon)}{\Gamma(1-2\epsilon)} (4\pi)^\epsilon 
       \left(\frac{1}{\epsilon}\right) \right. \nonumber \\ 
&   &\left. \times \int_{x_1}^{1-\delta_s}P_{bb}(z) 
     \frac{dz}{z} b(x_1/z, \mu_F)  +( b \leftrightarrow \bar{b} ) \right\} \nonumber \\
& & +(1 \leftrightarrow 2) \quad .
\label{ap}
\end{eqnarray}
The $1/\epsilon$ poles in the second line cancel the collinear 
singularities of the soft gluon corrections in Eq.~\ref{soft-1}, 
while the divergences in the third line cancel with the hard/collinear 
divergences in Eq.~\ref{hard/c}. 
These cancellations leave a finite cross section that has dependence 
on $\mu_R$ and $\mu_F$.

The remaining region has hard/non-collinear gluons and yields
a finite contribution to the cross section.
We compute this contribution numerically.

\subsection{Corrections from $bg \to b hh$}

Now let us consider the contributions from the parton subprocess
$b g \to b hh$, which is an ${\cal O}(1/\Lambda)$ correction 
to the LO process of $b\bar{b} \to hh$.
In calculating this cross section, 
again we ignore the bottom quark mass, $m_b $, except 
in the Yukawa couplings where the running mass is used. 
There are no IR singularities in the $b g \to b hh$ subprocess.
However, when we integrate over the momenta of the $b$ quarks, there
are initial state collinear singularities which arise from 
gluon splitting to a pair of collinear $b$ quarks. These singularities
are absorbed into gluon parton distribution functions.


\begin{figure}[htb]
\centering\leavevmode
\epsfxsize=4.5in
\epsfbox{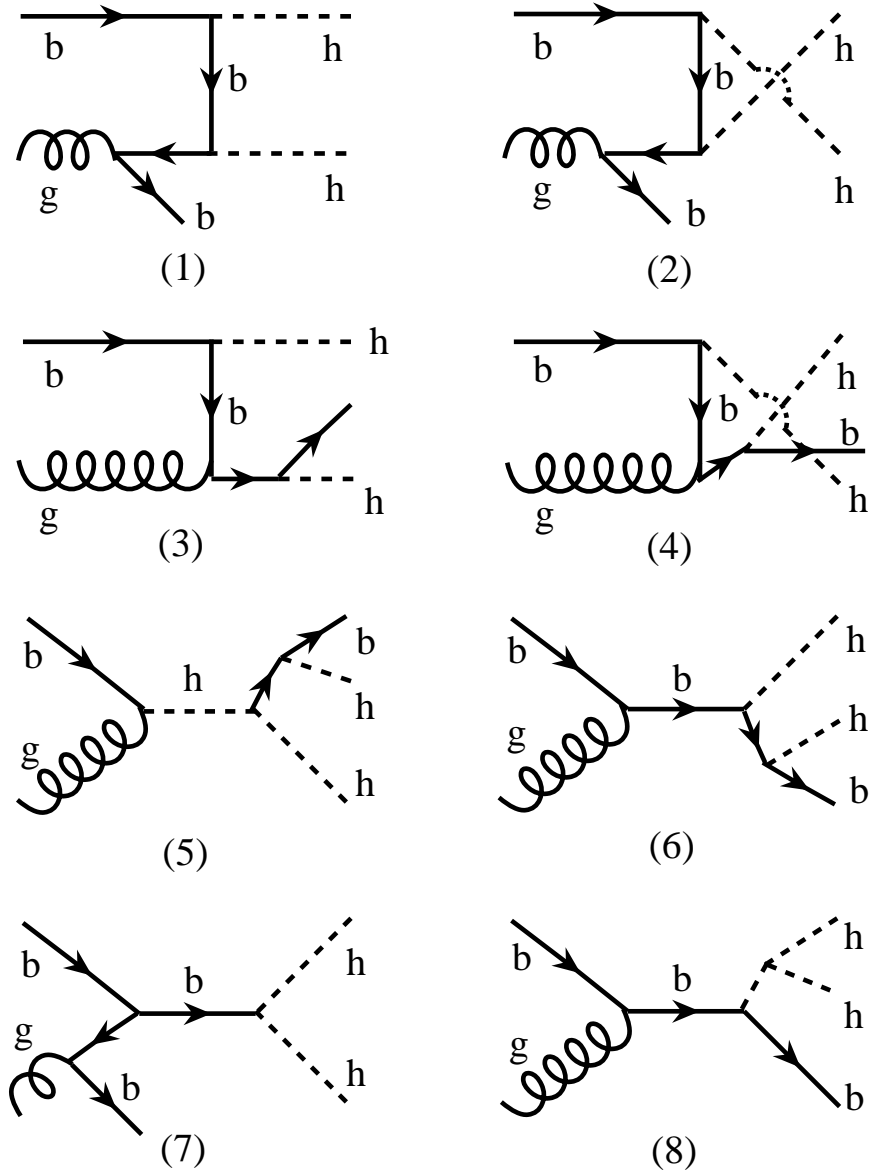}
\caption[]{The lowest order Feynman diagrams for $bg \to bhh$.}
\label{fig:bgtree}
\end{figure}

The eight diagrams for the $b g \to b hh$ subprocess are shown in 
Fig.~\ref{fig:bgtree}.  We assign momenta to partons as
$$b(p_1) g(p_2) \to b(p_b) h(p_3) h(p_4)\quad . $$

As there are no IR divergences, we do
not need to separate the final $b$ quark 
phase space into soft and hard regions. The cross section is:
\begin{eqnarray}
\sigma_{bg} & = &  \int dx_1 dx_2 
b(x_1) g(x_2) \hat{\sigma}(b g \to b hh)+ (1 \leftrightarrow 2)\quad .
\label{bg:lo}
\end{eqnarray}

As shown in diagrams (1), (2) and (7) of Fig.~\ref{fig:bgtree}, 
the initial gluon can be considered to split into two $b$ quarks. 
When these two $b$ quarks are parallel to each other, collinear 
singularities appear. To remove these collinear singularities, 
we introduce a collinear cutoff ($\delta_c$) to split the final 
$b$ quark phase space into collinear and non-collinear regions. 
In the  collinear region, the final $b$ is emitted within an angle 
satisfying,
\begin{eqnarray}
\frac{-(p_2 - p_b)^2}{E_b \sqrt{s}} < \delta_c \quad .
\end{eqnarray}

Using the same method as for the 
$b\bar{b}\to hhg$ real gluon emission corrections,
 we find the cross section in the collinear region:
\begin{eqnarray}
\sigma_{c} & = & \int dx_1 dx_2 \; 
 b(x_1) \hat{\sigma} \frac{\alpha_s}{2\pi}
\frac{\Gamma(1-\epsilon)}{\Gamma(1-2\epsilon)} 
\left(\frac{4\pi\mu^2}{s}\right)^\epsilon (-\frac{1}{\epsilon}) 
\delta_c^{-\epsilon}\nonumber \\
&& \times \int_{x_2}^{1} \frac{dz}{z} g(x_2/z) P_{bg}(z,\epsilon) 
          \left[\frac{(1-z)^2}{2z}\right]^{-\epsilon} \quad .
\label{h/col}
\end{eqnarray}
$P_{bg}$ is the Altarelli-Parisi splitting function 
for $g \rightarrow b b^\prime $ at lowest order,
\begin{eqnarray}
P_{bg}(z, \epsilon) & = & \frac{1}{2} \left[z^2 +(1-z)^2\right] -\epsilon z(1-z)\nonumber \\
&=& P_{bg}(z) + \epsilon P^\prime_{bg}(z)\quad .
\end{eqnarray}
 Again we introduce a modified parton distribution function,
\begin{eqnarray}
g(x, \mu_F)& = &g(x) 
 + \frac{\alpha_s}{2\pi}\frac{\Gamma(1-\epsilon)}
{\Gamma(1-2\epsilon)} \left( 4\pi \right)^\epsilon 
\left(-\frac{1}{\epsilon}\right) \int_{x}^{1}P_{bg}(z) \frac{dz}{z} g(x/z)
\label{g:pdf}
\end{eqnarray}
The contribution to the NLO total cross section from the $bg$ initial state
is then,
\begin{eqnarray}
\delta \sigma_{NLO}^{bg} & = &  \sigma_{c} + \sigma_{nc} \nonumber \\
& = & \frac{\alpha_s}{2\pi} \int dx_1 dx_2  \int_{x_2}^{1}
\frac{dz}{z} \hat{\sigma}_{LO} \; b(x_1,\mu_F) g(x_2/z,\mu_F)  \nonumber \\
& & \times \left[\frac{z^2+(1-z)^2}{2} \ln\left(\frac{s}{\mu^2}\frac{(1-z)^2}{z}\frac{\delta_c}{2}\right) +z(1 -z) \right]  \nonumber \\
&  & + \int dx_1 dx_2 \; b(x_1,\mu_F) g(x_2,\mu_F) 
\hat{\sigma}_{nc} (bg \to b hh) +( 1 \leftrightarrow 2 ) \quad .
\label{bg:sigma1}
\end{eqnarray}
Here $\hat{\sigma}_{LO}$ is the Born cross section for  
$b\bar{b} \rightarrow h h $ and 
$\hat{\sigma}_{nc}$ is the Born cross section for
 $b g \rightarrow b h h $.
 The $ \bar{b} g \rightarrow \bar{b} hh $ process has 
exactly the same contribution as $ b g \rightarrow bhh $.

\subsection{Next-to-leading order cross section for $b\bar{b} \to hh$}

We have obtained the virtual and the real gluon emission
${\cal O}(\alpha_s)$ corrections to 
$b\bar{b} \to h h$. The real gluon emission corrections 
include three regions: soft, hard/collinear and hard/non-collinear. 
The total next-to-leading order cross section can now be assembled
from the lowest order cross section, the virtual corrections, 
and the real gluon emission corrections. Summing all of the above
pieces, we obtain the ${\cal O}(\alpha_s)$
next-to-leading-order cross section  corrections,
\begin{eqnarray}
\sigma_{\rm Born} +\delta\sigma_{\alpha_s} 
& = & \sigma_{\rm Born}
     +\sigma_v +\sigma_{soft} +\sigma_{hard/c} +\sigma_{hard/nc} \nonumber \\
& = & \int dx_1 dx_2 \quad b(x_1,\mu) \bar{b}(x_2,\mu) 
      \left\{ \hat{\sigma}_{LO}
      \left[ 1 -\frac{4\alpha_s}{3\pi}\ln(\mu^2)\left(\ln(\delta_s^2)
             +\frac{3}{2}\right) \right] +\hat{\sigma}_{finite}\right\}  
      \nonumber \\
&   &+\frac{\alpha_s}{2\pi} C_F \int dx_1 dx_2 
       \int_{x_1}^{1-\delta_s} \frac{dz}{z} \hat{\sigma}_{LO} 
              \left[ b(x_1/z,\mu)\bar{b}(x_2,\mu) 
                    +b(x_2,\mu) \bar{b}(x_1/z,\mu) \right] 
      \nonumber \\
&   & \;\; \times \left[\frac{1+z^2}{1-z} \ln
     \left( \frac{s}{\mu^2}\frac{(1-z)^2}{z}\frac{\delta_c}{2}\right) 
     +1 -z \right] \nonumber \\
&  & +\int dx_1 dx_2  \; b(x_1,\mu) \bar{b}(x_2,\mu) 
      \hat{\sigma}_{hard/nc}(b\bar{b}\to hhg)\nonumber \\
&  & + (1 \leftrightarrow 2)
\label{hhg:sigma}
\end{eqnarray}
where $\hat{\sigma}_{finite}$ represents the finite cross section with
contributions from both virtual and soft gluon radiative corrections,
\begin{eqnarray}
\hat{\sigma}_{finite} & = & \int\frac{1}{2
  s}\frac{1}{2}\frac{4\alpha_s}{3\pi}\left\{ \left[ \frac{1}{2}
  \ln^2(s \delta_s^2) -\frac{\pi^2}{3}\right] |M_0|^2  
+|M_D|^2\right\} d\Phi_2(b\bar{b} \to hh) \; .
\end{eqnarray}
We have taken $\mu=\mu_R=\mu_F$.

The ${\cal O}(1/\Lambda)$ corrections from the subprocesses
$bg \to b hh$ and $ \bar{b} g \rightarrow \bar{b} hh $ include
collinear and non-collinear contributions and yield the $1/\Lambda$ 
corrections for Higgs pair production associated with one $b$ quark. 
\begin{eqnarray}
\delta \sigma_{1/\Lambda} 
& = & \delta \sigma_{NLO}^{bg} + \delta \sigma_{NLO}^{\bar{b}g} \nonumber \\
\label{bg:sigma}
\end{eqnarray}
Summing Eq.~\ref{hhg:sigma} and ~\ref{bg:sigma}, 
we get the total next-to-leading-order cross section for
Higgs pair production from bottom quark fusion,
\begin{eqnarray}
\sigma_{NLO} 
& = & \sigma_{\rm Born}
     +\delta\sigma_{\alpha_s} +\delta\sigma_{1/\Lambda} \; .
\end{eqnarray}

The associated Higgs boson pair production with $b\bar{b}$ occurs via 
gluon fusion $gg \to b\bar{b} hh$ and quark-antiquark annihilation 
$q\bar{q} \to b\bar{b} hh$. 
These are subleading corrections to the NLO results given above. 
To estimate the cross section from these subprocesses, we have
applied a nonzero value of $m_b = 4.7$ GeV for the bottom quark mass except 
in the Yukawa couplings where the running mass is used. 
We use MadGraph~\cite{MadGraph} and HELAS~\cite{HELAS} to calculate 
the cross sections for these processes and find that the contribution 
from $q\bar{q} \to b\bar{b} hh$ is much smaller than that from 
$gg \to b\bar{b} hh$.

\section{Results for Higgs pair production in bottom quark fusion}

In this section, we present our results for the next-to-leading-order 
inclusive cross section for $p p \to h h +X$ via bottom quark fusion, 
$b\bar{b} \to h h $.
We use the lowest order CTEQ6L1 parton distribution functions 
(PDFs)~\cite{Pumplin:2002vw} at the factorization scale $\mu_F$ 
with the leading-order evolution of the strong coupling
$\alpha_s(\mu_R)$ at the renormalization scale $\mu_R$ 
to calculate the LO cross section and 
the CTEQ6M PDFs at $\mu_F$ with the next-to-leading-order evolution of 
$\alpha_s(\mu_R)$ to evaluate the NLO inclusive cross section.

 
\begin{figure}[htb]
\centering\leavevmode
\epsfxsize=6in
\epsfbox{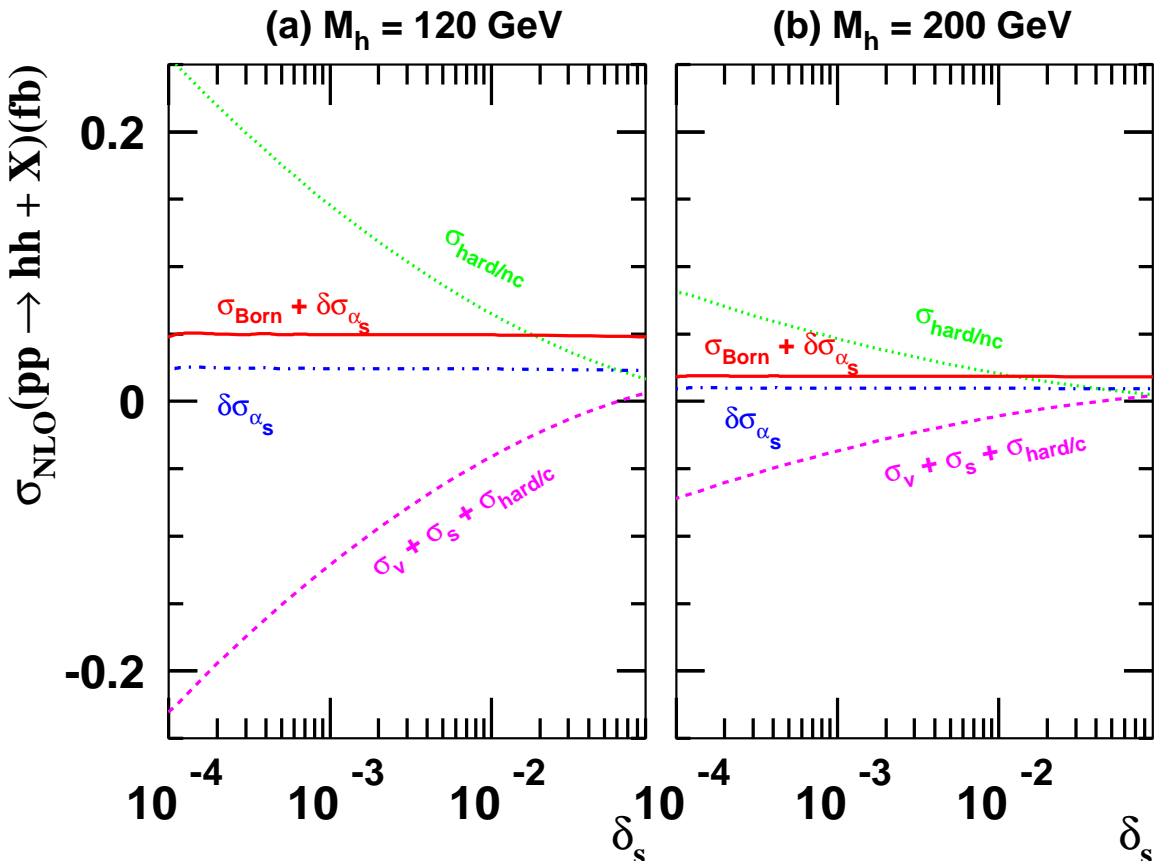}
\caption[]{
Order $\alpha_s$ corrections $\delta \sigma_{\alpha_s}(b\bar{b} \to hh)$ 
(dash-dot, blue) versus the soft cutoff $\delta_s$ with $\sqrt{S}= 14$~TeV, 
$\mu_R = \mu_F = M_h/2$ and $\delta_c = \delta_s/10$ for
(a) $M_h = 120$~GeV and (b) $M_h = 200$~GeV. 
These corrections have contributions from 
virtual gluons ($\sigma_v$), soft gluon emission ($\sigma_{s}$), 
hard/collinear gluon emission ($\sigma_{hard/c}$),
and hard/non-collinear gluon emission ($\sigma_{hard/nc}$).
These graphs show the cancellation of the $\delta_s$ dependence between 
$\sigma_v + \sigma_{s} + \sigma_{hard/c}$ (dash, magenta) 
and $\sigma_{hard/nc}$ (dot, green). 
Also shown is the sum of $\sigma_{\rm Born}$ and 
$\delta\sigma_{\alpha_s}$ (solid, red). 
}
\label{fig:delts}
\end{figure}


\begin{figure}[htb]
\centering\leavevmode
\epsfxsize=6in
\epsfbox{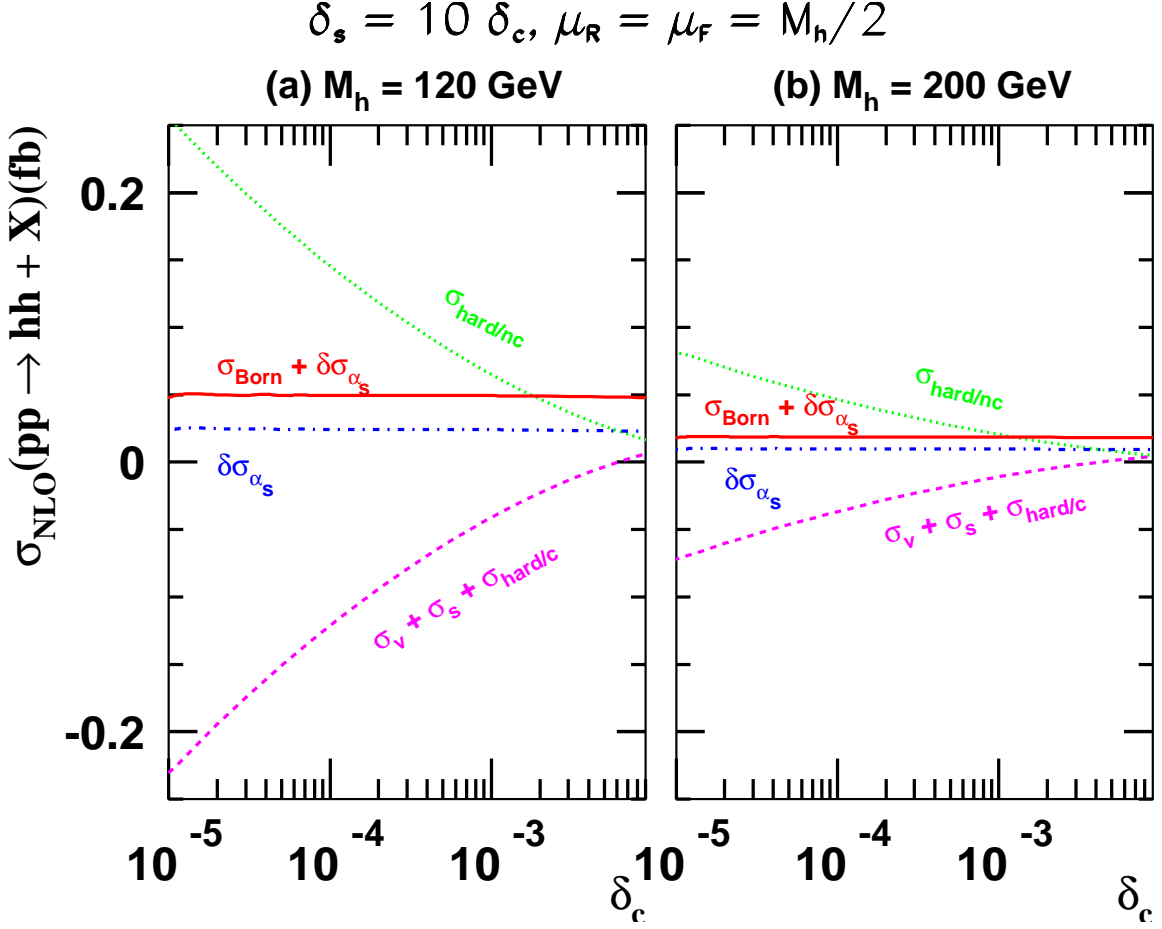}
\caption[]{
Order $\alpha_s$ corrections $\delta \sigma_{\alpha_s}(b\bar{b} \to hh)$ 
(dash-dot, blue) versus the hard/collinear cutoff $\delta_c$ 
with $\sqrt{S}= 14$~TeV, $\mu_R = \mu_F = M_h/2$ and 
$\delta_s = 10 \delta_c$ for (a) $M_h = 120$~GeV and (b) $M_h = 200$~GeV. 
These corrections have contributions from 
virtual gluons ($\sigma_v$), soft gluon emission ($\sigma_{s}$), 
hard/collinear gluon emission ($\sigma_{hard/c}$),
and hard/non-collinear gluon emission ($\sigma_{hard/nc}$).
These graphs show the cancellation of the $\delta_c$ dependence between 
$\sigma_v + \sigma_{s} + \sigma_{hard/c}$ (dash, magenta) 
and $\sigma_{hard/nc}$ (dot, green). 
Also shown is the sum of $\sigma_{\rm Born}$ and 
$\delta\sigma_{\alpha_s}$ (solid, red). 
}
\label{fig:deltc}
\end{figure}

We have introduced two arbitrary small cutoffs, the soft cutoff $\delta_s$ 
and the collinear cutoff $\delta_c$, to split the phase space in the
real gluon emission corrections into soft, hard/collinear and 
hard/non-collinear regions. These separations are not physical and 
our final results are reliable only if they are not sensitive to 
these parameters. 

To check the dependence of the ${\cal O}(\alpha_s)$ NLO cross section 
on these two parameters, we present the order $\alpha_s$ corrections 
for Higgs pair production 
$\delta\sigma_{\alpha_s}(b\bar{b} \to hh)$ versus $\delta_s$ 
in Fig.~\ref{fig:delts} and 
$\delta\sigma_{\alpha_s}(b\bar{b} \to hh)$ versus $\delta_c$ 
in Fig.~\ref{fig:deltc} 
with $\sqrt{S}= 14$~TeV for (a) $M_h = 120$~GeV and (b) $M_h = 200$~GeV. 
These corrections have contributions from 
virtual gluons ($\sigma_v$), soft gluon emission ($\sigma_{s}$), 
hard/collinear gluon emission ($\sigma_{hard/c}$),
and hard/non-collinear gluon emission ($\sigma_{hard/nc}$).
These graphs show the cancellation of the $\delta_s$ dependence and 
the $\delta_c$ dependence between 
$\sigma_v + \sigma_{s} + \sigma_{hard/c}$ and $\sigma_{hard/nc}$.
Also shown is the sum of $\sigma_{\rm Born}$ and $\delta\sigma_{\alpha_s}$.
We have chosen $\delta_c = \delta_s/10$ and 
the renormalization/factorization scales to be $\mu_F = \mu_R = M_h/2$.

There are several points to note. 
\begin{itemize}
\item We divide the phase space of the real gluon emission correction into 
soft, hard/collinear and hard/non-collinear regions by 
introducing two small cut-offs $\delta_s$ and $\delta_c$. The correction 
in each region is then integrated over the corresponding phase space. 
As we can see in these graphs, the corrections in each region are 
very sensitive to the values of $\delta_s$ and $\delta_c$.
\item Since these two small cutoffs, $\delta_s$ and $\delta_c$, are arbitrary,
the total cross section should not depend on either one of them. 
These two figures show the cancellation of the $\delta_s$ and
$\delta_c$ dependences. 
\item At the renormalization/factorization scale $\mu_R = \mu_F = M_h/2$, 
the ${\cal O}(\alpha_s)$ NLO cross section corrections are comparable 
with the Born cross section. This implies that the ${\cal O}(\alpha_s)$ 
corrections significantly increase the LO cross section.
\end{itemize}


\begin{figure}[htb]
\centering\leavevmode
\epsfxsize=6in
\epsfbox{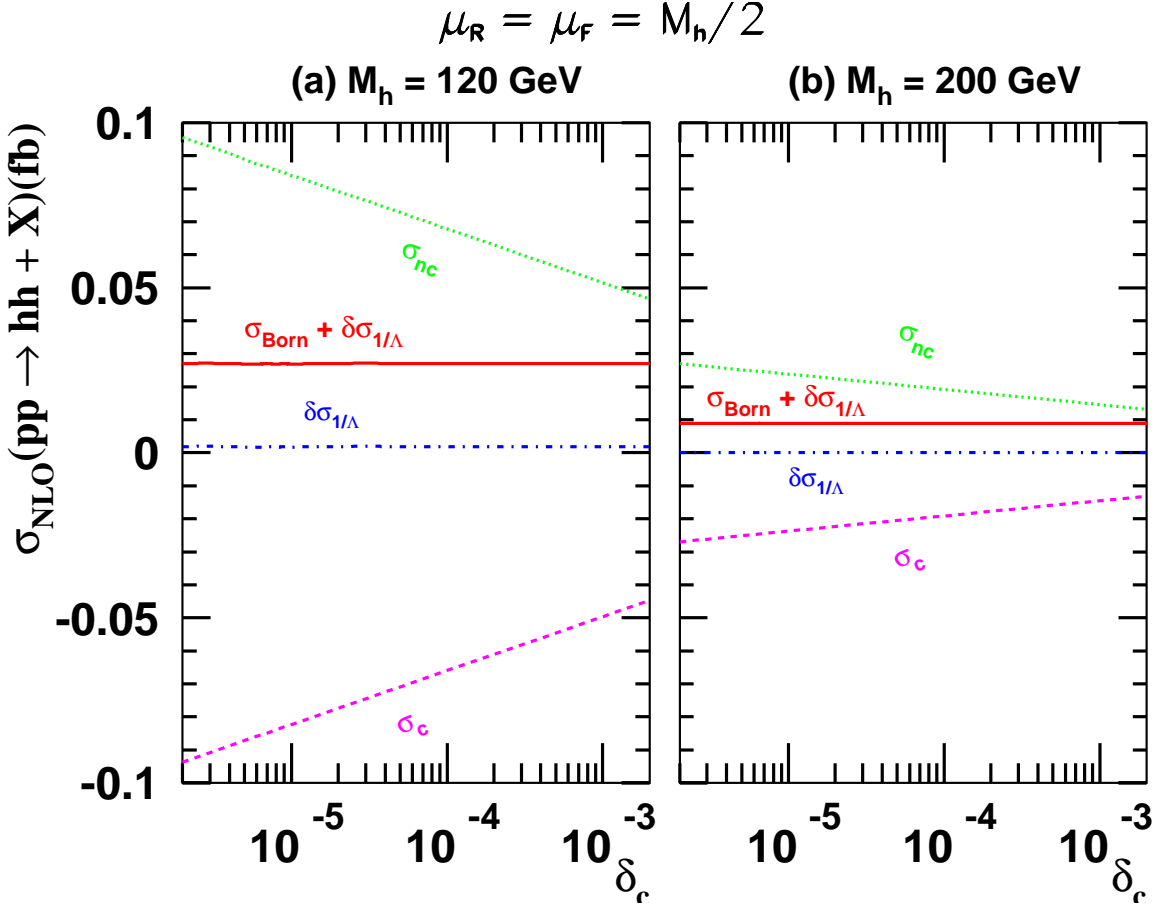}
\caption[]{
Order $1/\Lambda$ corrections $\delta \sigma(1/\Lambda)(bg \to bhh)$ 
(dash-dot, blue) versus the collinear cutoff $\delta_c$ with 
$\sqrt{S} = 14$~TeV and $\mu_R = \mu_F = M_h/2$ for (a) $M_h = 120$~GeV 
and (b) $M_h = 200$~GeV.  
This figure shows the cancellation of the $\delta_c$ dependence between 
$\sigma_c$ (dash, magenta) and $\sigma_{nc}$ (dot, green).
Also shown is the sum of $\sigma_{\rm Born}$ and
$\delta\sigma_{1/\Lambda}$ (solid, red).
}
\label{fig:bg_deltc}
\end{figure}

Figure~\ref{fig:bg_deltc} shows the independence of order $1/\Lambda$ 
corrections ($\delta\sigma_{1/\Lambda}$) on the collinear cutoff $\delta_c$ 
in the $bg \to bhh$ subprocess. 
Like the $\alpha_s$ corrections to $b\bar{b} \to hh$, the contributions from 
collinear and non-collinear regions are very sensitive to the 
values of $\delta_c$, but the total ${\cal O}(1/\Lambda)$  
NLO cross section correction is independent of $\delta_c$. 
The ${\cal O}(1/\Lambda)$ contributions from 
 $bg \to bhh$ are much smaller than the 
${\cal O}(\alpha_s)$ NLO corrections from the $b\bar{b}$ initial state. 

\begin{figure}[htb]
\centering\leavevmode
\epsfxsize=6in
\epsfbox{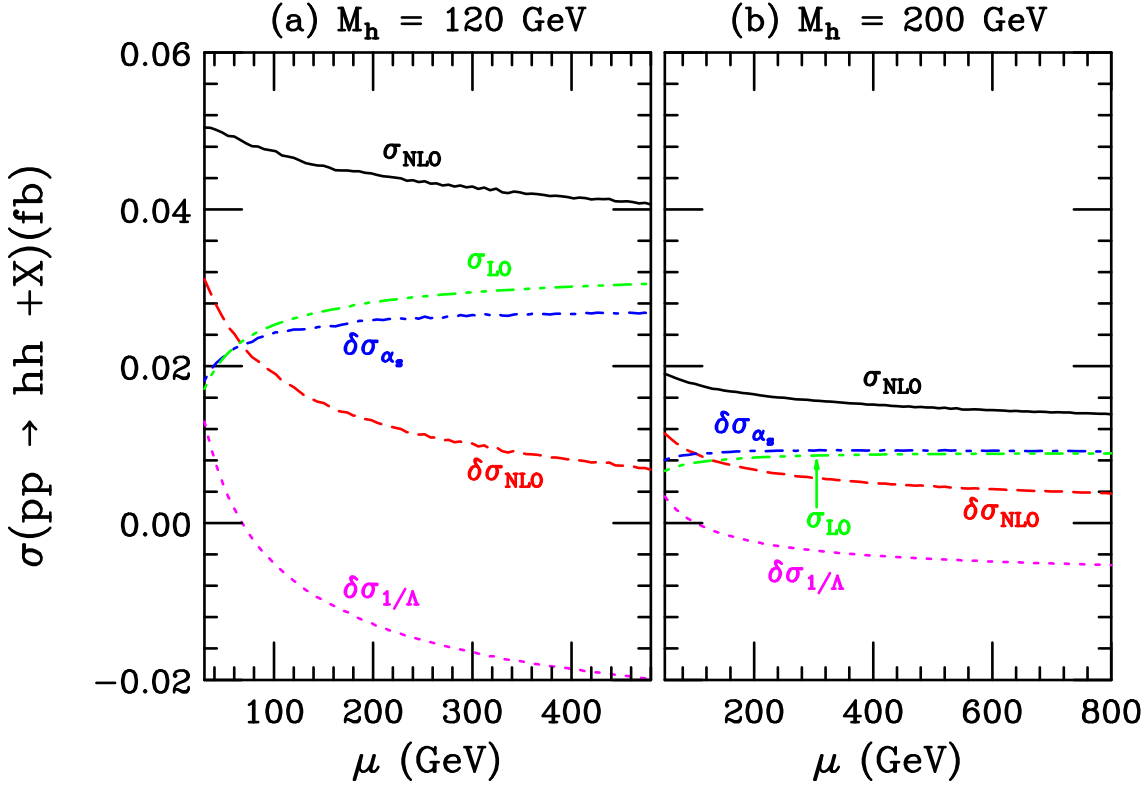}
\caption[]{Next-to-leading order cross section 
$\sigma_{NLO}(pp \to hh +X)$ (solid, black) in bottom quark fusion versus 
renormalization/factorization scale $\mu = \mu_R = \mu_F$ 
with $\sqrt{S}= 14$~TeV, $\delta_s = 10^{-3}$ and $\delta_c = 10^{-4}$, 
for (a) $M_h = 120$~GeV and (b) $M_h = 200 $~GeV. 
Also shown are the LO cross section $\sigma_{\rm LO}$ (dash-dot-dot, green), 
$\alpha_s$ corrections $\delta\sigma_{\alpha_s}$ (dash-dot, blue), 
$1/\Lambda$ corrections $\delta\sigma_{1/\Lambda}$ (dot, magenta), 
and the NLO correction 
$\delta\sigma_{\rm NLO} = \delta\sigma_{\alpha_s} 
+\delta\sigma_{1/\Lambda}$ (dash, red).
}
\label{fig:mu}
\end{figure}

In Fig.~\ref{fig:mu}, we study the dependence of the LO and NLO cross
sections on the renormalization and factorization scales. 
We present the next-to-leading order cross section 
$\sigma_{\rm NLO}(pp \to hh +X)$ via bottom quark fusion versus 
$\mu = \mu_R = \mu_F$ with $\delta_s = 10^{-3}$ and $\delta_c = 10^{-4}$ 
for (a) $M_h = 120 $ GeV and (b) $M_h = 200 $ GeV.
Also shown are the LO cross section ($\sigma_{\rm LO}$),
$\alpha_s$ corrections ($\delta\sigma_{\alpha_s}$), 
$1/\Lambda$ corrections ($\delta\sigma_{1/\Lambda}$),
and the NLO correction 
$\delta\sigma_{\rm NLO} = \delta\sigma_{\alpha_s}
+\delta\sigma_{1/\Lambda}$.

We note that:
\begin{itemize}
\item The total NLO cross section corrections decrease with $\mu$. 
The larger the Higgs mass, the slower the decrease of the NLO corrections.
\item
The NLO cross section has less dependence on the renormalization 
and factorization scales than does the LO cross section.
\end{itemize}


\begin{figure}[htb]
\centering\leavevmode
\epsfxsize=6in
\epsfbox{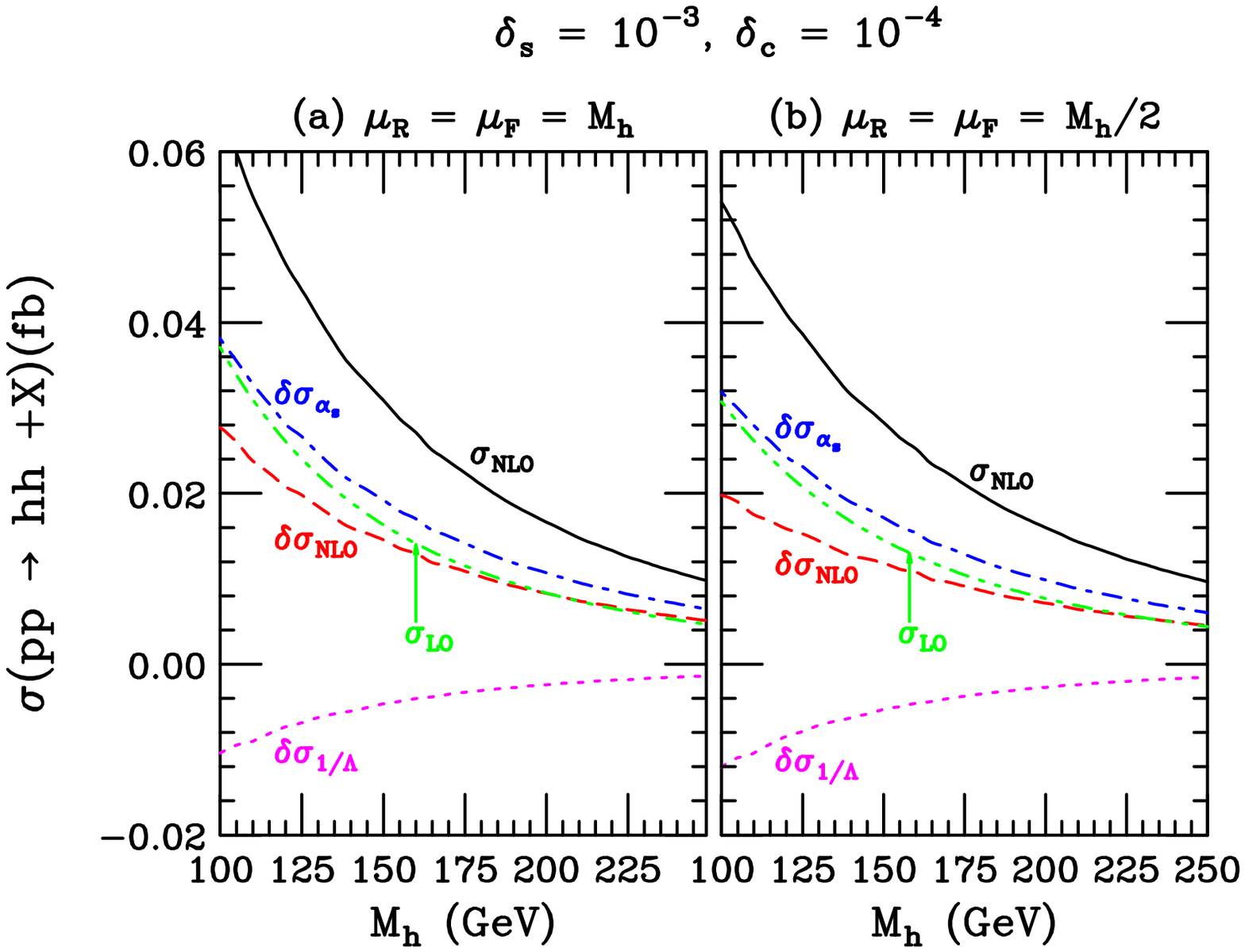}
\caption[]{Next-to-leading order cross section 
$\sigma_{NLO}(pp \to hh +X)$ (solid, black) in bottom quark fusion versus 
the Higgs mass ($M_h$) with $\sqrt{S}= 14$~TeV, $\delta_s = 10^{-3}$
and $\delta_c = 10^{-4}$, for (a) $\mu_R = \mu_F = M_h$ and 
(b) $\mu_R = \mu_F = M_h/2$.
Also shown are the LO cross section $\sigma_{\rm LO}$ (dash-dot-dot, green), 
$\alpha_s$ corrections $\delta\sigma_{\alpha_s}$ (dash-dot, blue), 
$1/\Lambda$ corrections $\delta\sigma_{1/\Lambda}$ (dot, magenta), 
and the NLO correction 
$\delta\sigma_{\rm NLO} = \delta\sigma_{\alpha_s} 
+\delta\sigma_{1/\Lambda}$ (dash, red).
}
\label{fig:higgs}
\end{figure}

Fig.~\ref{fig:higgs} shows the NLO total cross section $pp \to hh +X$ 
in bottom quark fusion as a function of the Higgs mass ($M_h$) at the LHC 
with $\sqrt{S} = 14$~TeV, $\delta_s = 10^{-3} $ and $\delta_c = 10^{-4}$.
The renormalization and factorization scales are chosen to be 
(a) $\mu = M_h$ and (b) $\mu = M_h/2$.  The NLO cross section 
correction for $\mu = M_h$ is bigger than for $\mu = M_h/2$. 
In addition, we present the LO cross section ($\sigma_{\rm LO}$),
$\alpha_s$ corrections ($\delta\sigma_{\alpha_s}$), 
$1/\Lambda$ corrections ($\delta\sigma_{1/\Lambda}$),
and the NLO correction 
$\delta\sigma_{\rm NLO} = \delta\sigma_{\alpha_s}
+\delta\sigma_{1/\Lambda}$.


\begin{figure}[htb]
\centering\leavevmode
\epsfxsize=6in
\epsfbox{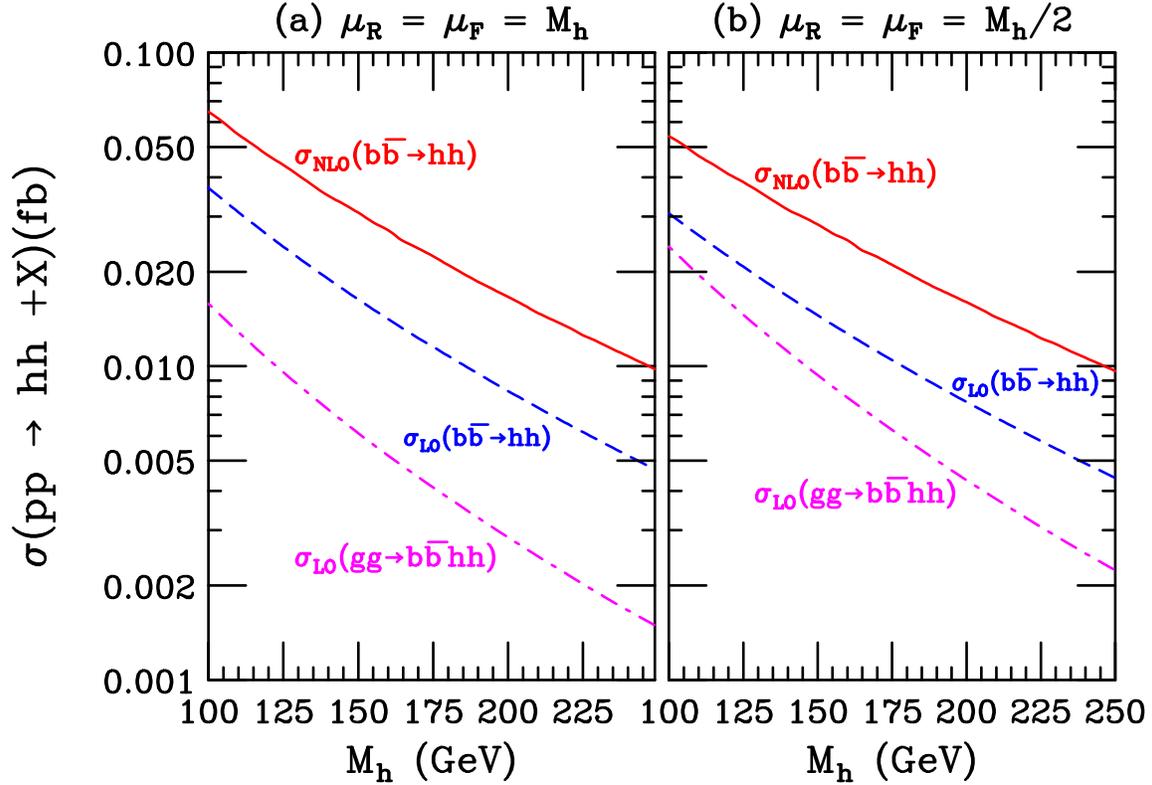}
\caption[]{Leading order cross section $\sigma_{LO}$ (dash, blue) and 
next-to-leading order cross section $\sigma_{NLO}$ (solid, red) 
for $pp \to hh +X$ via bottom quark fusion as a function of $M_h$ 
at the LHC with $\sqrt{S} = 14$ TeV 
for (a) $\mu = M_h$ and (b) $\mu = M_h/2$. 
Also shown is the ${\cal O}(1/\Lambda^2)$ contribution from 
$gg \to b\bar{b} hh$.
}
\label{fig:sigma}
\end{figure}

In Fig.~\ref{fig:sigma}, we show the LO and the NLO cross section
including the ${\cal O}(\alpha_s)$ contributions from the
$b {\overline b}$ initial state and the ${\cal O}(1/\Lambda)$ contribution
from the $bg$ initial state. We separately show the  ${\cal O}(1/\Lambda^2)$ 
contribution via $gg \to b\bar{b}hh$ as a function of the Higgs mass ($M_h$) 
at the LHC with $\sqrt{s} = 14$~TeV. 
As in Fig.~\ref{fig:higgs}, we present our results at two 
renormalization/factorization scales, 
(a) $\mu = M_h$ and (b) $\mu = M_h/2$ 
with $\delta_s = 10^{-3} $ and $\delta_c = 10^{-4}$. 

In calculating the sub-leading contribution from the subprocess
$gg \to b\bar{b}hh$, we have evaluated the cross section numerically for
$pp \to b\bar{b} hh +X$ via $gg \to b\bar{b} hh$ 
with a finite quark mass to regulate the collinear singularity.
We take the $b$ quark mass to be $m_b =$ 4.7 GeV.

As is shown in Fig.~\ref{fig:sigma}, both the LO and NLO 
total cross section 
for the  $bb \to hh $ process at a renormalization/factorization
scale $M_h$  are bigger than the
corresponding cross sections at the scale $M_h/2$.


In the Standard Model, gluon fusion is the dominant source for
producing a pair of Higgs bosons through triangle and box diagrams 
containing top and bottom quark loops.
It has been demonstrated that it might be possible to study 
the trilinear Higgs coupling at the LHC through the gluon fusion
production mechanism of Higgs boson pairs~\cite{
Boudjema:1995cb,Djouadi:1999rc,Muhlleitner:2003me,Baur:2003gp,Moretti:2004wa}.
In the minimal supersymmetric standard model, bottom quark fusion 
could offer great promise to study the trilinear couplings of Higgs
bosons for  large values of $\tan\beta$ through $b{\overline b} \rightarrow
hh$.

\begin{figure}[htb]
\centering\leavevmode
\epsfxsize=6in
\epsfbox{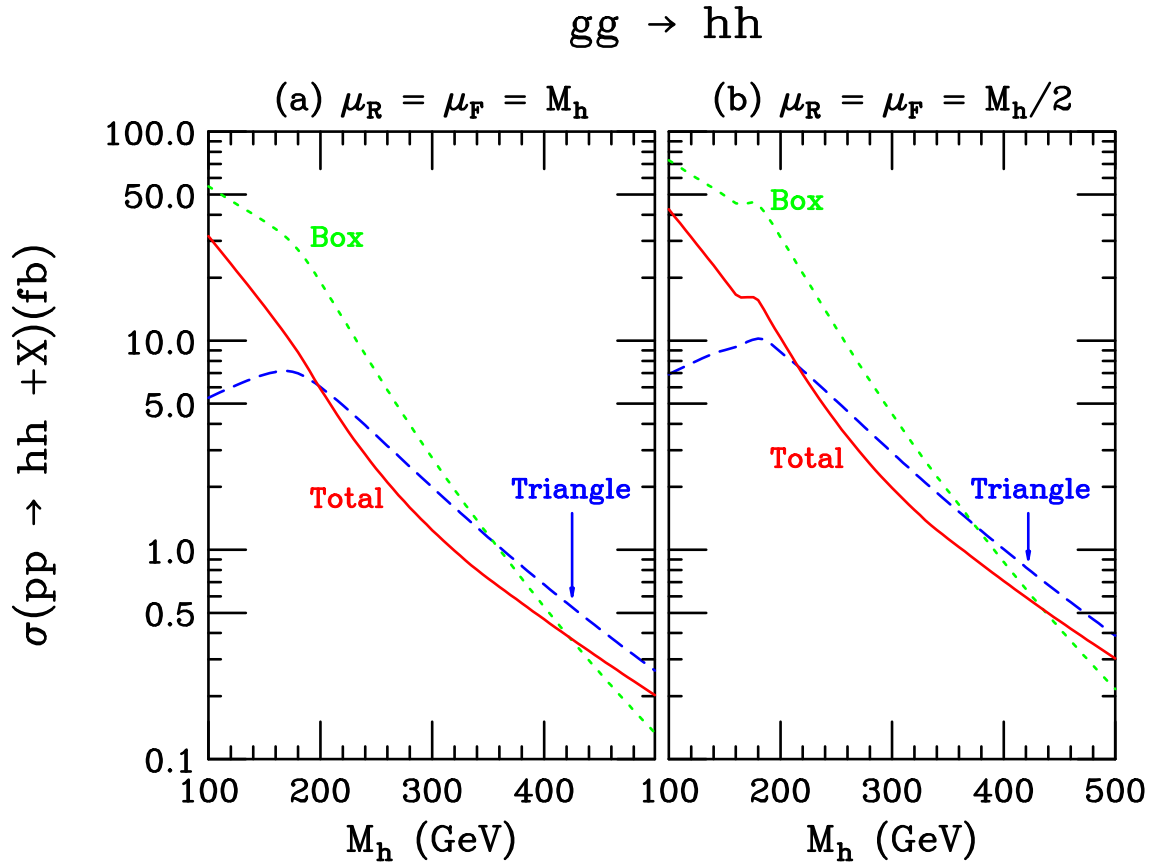}
\caption[]{Cross section of $pp \to hh +X$ via gluon fusion versus
the Higgs mass ($M_h$) at LHC with $\sqrt{S} = 14$ TeV for the
contributions from the triangle diagrams (dash, blue) 
and the box diagrams (dot, green) respectively. 
Also shown is the total cross section (solid).
The renormalization and factorization scales are chosen to be 
(a) $\mu_R = \mu_F = M_h$ and (b) $\mu_R = \mu_F = M_h/2$. 
}
\label{fig:gghh}
\end{figure}

In Figure~\ref{fig:gghh}, we show the cross 
section for $pp \to hh +X$ via gluon
fusion ($gg \to hh$) through top and bottom loop diagrams as a
function of $M_h$ at LHC with $\sqrt{S} = 14$ TeV for (a) $\mu = M_h$ 
and (b) $\mu = M_h/2$.  In the SM this production mechanism is 
significantly larger than the $b {\overline b}$ fusion mechanism. 


\section{Conclusions}

In this paper we present complete ${\cal O}(\alpha_s)$ and 
${\cal O}(1/\Lambda)$ QCD corrections 
to the production of a pair of Higgs bosons via bottom quark fusion at 
the LHC. We introduce two arbitrary small cutoffs, 
$\delta_s$ and $\delta_c$, to compute ${\cal O}(\alpha_s)$
NLO soft, hard/collinear and 
hard/non-collinear gluon emission corrections to  $b\bar{b} \to hh$. 
The total ${\cal O}(\alpha_s)$ 
NLO cross section corrections are independent of the values
of $\delta_s$ and $\delta_c$.

Our results are not sensitive to the difference between renormalization and 
factorization scales and we use the same renormalization and
factorization scales. 
The LO cross section is very sensitive to the factorization scale. 
The cross section for the $b\bar{b} \to hh$ process with large
 factorization
scale is larger than the corresponding cross section  with 
a low factorization scale.

The rate for double Higgs production in the SM is very small, although
the NLO corrections we have computed significantly increase this rate. 
However, the rate for Higgs pair production will be enhanced in models 
with large couplings of the Higgs bosons to $b$ quarks.  
Our results are of interest in attempts to measure the trilinear 
Higgs coupling in such models.

\section*{Acknowledgments}

We are grateful to Jeff Owens, Scott Willenbrock and Jianwei Qiu 
for beneficial discussions. 
C.K. thanks the Theoretical Physics Department of Fermilab for 
hospitality and support during a sabbatical visit.
This research was supported in part by the U.S. Department of Energy
under Grants No.~DE-AC02-76CH1-886,
No.~DE-FG02-04ER41305 and No.~DE-FG02-03ER46040.

\appendix

\section{Heavy Quark Running Mass}

The two-loop running mass of a heavy quark in the $\overline{MS}$ scheme has the expression~\cite{Dicus:1998hs,Vermaseren:1997fq}

\begin{equation}
\bar{m}(\mu) = \bar{m}(\mu_0)\left(\frac{\alpha_s(\mu)}{\alpha_s(\mu_0)}\right)^{\gamma_0/\beta_0}\frac{\left[1+ a_1 \frac{\alpha_s(\mu)}{\pi}  + (a_1^2 + a_2)/2 \left(\frac{\alpha_s(\mu)}{\pi}\right)^2\right]}{\left[1+ a_1 \frac{\alpha_s(\mu)}{\pi}  + (a_1^2 + a_2)/2 \left(\frac{\alpha_s(\mu)}{\pi}\right)^2\right]}
\end{equation}
where

\begin{eqnarray}
a_1 & = &-\frac{\beta_1 \gamma_0}{\beta_0^2 } + \frac{\gamma_1}{\beta_0} \nonumber \\
a_2 & = &\frac{\gamma_0}{\beta_0^2}\left(\frac{\beta_1^2}{\beta_0}-\beta_2\right)-\frac{\beta_1 \gamma_1}{\beta_0^2 } + \frac{\gamma_2}{\beta_0} \nonumber \\
\gamma_0 & = & 1 \nonumber \\
\gamma_1 & = & \frac{1}{16}\left(\frac{202}{3} - \frac{20}{9} N_f\right) \nonumber \\
\gamma_2 & = & \frac{1}{64}\left[1249 + \left( - \frac{2216}{27} - \frac{160}{3}\zeta_3\right)N_f - \frac{140}{81}N_f^2\right] \nonumber \\
\beta_0 & = & \frac{1}{4} \left(11 - \frac{2}{3} N_f\right) \nonumber \\
\beta_1 & = & \frac{1}{16} \left(102 - \frac{38}{3} N_f\right)
\end{eqnarray}

\noindent
where $N_f$ is the number of quark flavors with mass less than $\mu
(\mu > \mu_0)$, and $\zeta$ is the Riemann zeta function with 
$\zeta_3 = 1.2020569$.

The $\overline{MS}$ mass at next-leading order is
\begin{equation}
M_Q = \bar{m}(M_Q)\left( 1 + C_F \frac{\alpha_s}{\pi}\right)
\end{equation}

Here $M_Q$ is the  heavy quark pole mass. The evolution of the  strong
coupling can be found in Ref.~\cite{Marciano:1983pj}.
  

\end{document}